\documentclass[review]{elsarticle}
\usepackage{epsfig}
\usepackage{a4wide}
\usepackage{graphicx,amssymb}
\usepackage{amsmath
}        

\usepackage{color}
\usepackage{ulem}

\def\lb#1{\if 1#1 \ln\beta \else \ln^#1\beta \fi}
\def\lt#1{\if 1#1 \ln 2 \else \ln^#1 2 \fi}

\newcommand{\be}{\begin{equation}}
\newcommand{\ee}{\end{equation}}
\newcommand{\ba}{\begin{eqnarray}}
\newcommand{\ea}{\end{eqnarray}}

\newcommand{\ep}{\epsilon}

\newcommand{\lmmh}{\ln \frac{\mu^2}{m_H^2}}

\allowdisplaybreaks

\begin{document}

\vspace{\baselineskip}

\title{
\mbox{}\hfill {\normalsize SFB/CPP-13-29, TTP13-13, LPN13-030}
\\[2em]
On the Higgs boson pair  production at the LHC
}

\author[jg]{Jonathan Grigo}
\author[jg]{Jens Hoff}
\author[km]{Kirill~Melnikov} 
\author[jg]{Matthias Steinhauser}

\address[jg]{Institut f\"ur Theoretische Teilchenphysik, 
Karlsruhe Institute of Technology (KIT), Karlsruhe, Germany}

\address[km]{Department of Physics and Astronomy,
Johns Hopkins University,
Baltimore, MD, USA}

\begin{abstract}
  \noindent
  We compute the production cross section of a pair of Standard Model Higgs
  bosons at the LHC at next-to-leading order in QCD, including corrections in
  inverse powers of the top quark mass.  We calculate these power corrections
  through ${\cal O}(1/M_t^8)$ and study their relevance for phenomenology of
  the double Higgs production.  We find that power corrections are
  significant, even for moderate values of partonic center-of-mass energies,
  and that convergence of the $1/M_t$ expansion can be dramatically improved
  by factorizing the leading order cross section with full $M_t$-dependence.
\end{abstract}
      
\maketitle

\section{Introduction} 
The recent discovery \cite{atlasd, cmsd} of a scalar particle with properties
that are very similar to that of a Higgs boson completes the first stage of
the quest to understand the mechanism of electroweak symmetry breaking.
Indeed, after many years of preparation, ATLAS and CMS collaborations have
identified an elementary particle that may be directly connected to this
phenomenon.  Understanding this connection will be the primary focus of
particle physics in the coming years and there are two things that have to be
done. First, it is important to measure couplings of the Higgs boson  to gauge
bosons and fermions, and to verify that the Higgs couplings to various particles
are proportional to their masses.  Early measurements of the couplings
strengths seem to confirm this hypothesis \cite{atlasc}.  Second, it is
crucial to probe the Higgs boson self-interactions. Indeed, within the
Standard Model and many of its extensions, the Higgs boson self-interaction
triggers the electroweak symmetry breaking and it is important to verify that
we properly understand this phenomenon.

\begin{figure}[t]
  \begin{center}
    \includegraphics[width=0.95\columnwidth]{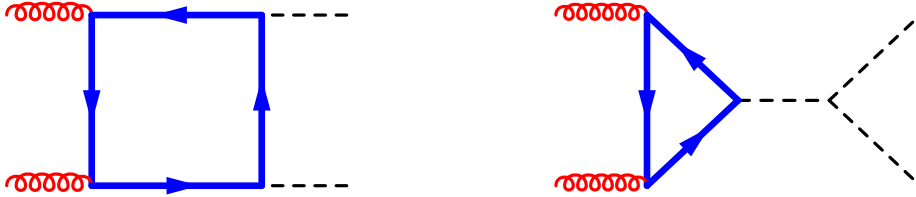}
    \caption{Box and triangle diagrams that contribute to double Higgs boson
      production at leading order.  Solid lines refer to top quarks and
      dashed lines refer to Higgs bosons.  }
    \label{fig0}
  \end{center}
\end{figure}

Self-interactions of the Higgs field and the electroweak symmetry breaking
induce the triple Higgs boson coupling. This coupling can be studied in the
production of a pair of Higgs bosons at the LHC \cite{bij,plehn,dj}.  At
leading order in QCD perturbation theory, the production of the Higgs boson
occurs in gluon fusion and proceeds either through a box $gg \to HH$ or a
triangle $gg \to H^*$ diagram, see Fig.~\ref{fig0}.  In the latter case the
off-shell Higgs boson decays to two Higgs bosons in the final state $H^* \to
HH$, making this contribution sensitive to triple Higgs boson coupling.
Unfortunately, observation of this process at the LHC is very challenging.
Indeed, it has a relatively small cross section to begin with and,
furthermore, it suffers from huge backgrounds that are present for almost all
major decay modes of the Higgs bosons.  Earlier estimates suggest that, with
$600~{\rm fb}^{-1}$, it is possible to study the double Higgs boson production
at the LHC in the $b \bar b \gamma \gamma$ channel \cite{baur}.  More recent
applications of jet substructure techniques to double Higgs production
indicate that one is sensitive to this process in $b \bar b W^+W^-$ and $ b
\bar b \tau \bar \tau $ channels with the integrated luminosity of $600
-1000~{\rm fb}^{-1}$ \cite{papa,dolan}.  Since substructure techniques are
still in the process of being developed, it is reasonable to expect
significant improvements in these initial estimates.  Therefore, we will
assume an optimistic outlook about prospects for measuring the double Higgs
boson production cross section at the LHC and we will try to improve the
quality of theoretical description of this process within the Standard Model.

We begin by summarizing what is known about the double Higgs boson production
in hadron collisions in the Standard Model.  As we already mentioned, the
dominant contribution to the process $pp \to HH$ is the gluon fusion that only
occurs at one-loop in perturbative QCD.  The corresponding partonic and
hadronic cross sections were computed in Refs.~\cite{bij,plehn}.  Part of this
contribution comes from $gg \to H^*$ process (see Fig.~\ref{fig0}) which is
equivalent to single Higgs boson production.  It is well-known that QCD
radiative corrections to single Higgs production are large
\cite{dawson,spira1,hnloexact}.  The bulk of these large radiative corrections
comes from relatively soft gluons that should not be sensitive to the details
of the final state as long as it is colorless. Therefore, we can expect that
similar large corrections are present in case of the double Higgs boson
production but, unfortunately, it is difficult to make this statement
precise. This is so because computation of QCD corrections to double Higgs
boson production requires four-point two-loop amplitudes with massive
particles which are out of reach of contemporary technology for perturbative
QCD computations.

Given this difficulty, computation of QCD corrections to double Higgs boson
production was performed in the approximation where the mass of the top quark
was taken to be very large compared to the partonic center-of-mass collision
energy and the Higgs boson mass \cite{Dawson:1998py}.  QCD corrections
obtained in this large-$M_t$ approximation appear to be significant; the ratio
of next-to-leading order cross section to leading-order cross section is close
to a factor two if factorization and renormalization scales are set to
$m_H$. However, it is unclear to what extent these large radiative corrections
are relevant for phenomenology because, as was repeatedly emphasized in the
literature \cite{Dawson:1998py,Dawson2013}, the large-$M_t$ approximation
fails to provide a good description of the $pp \to HH$ cross section for
realistic values of $M_t, m_H$ and the hadronic center-of-mass collision
energy.  To illustrate this point, we note that the large-$M_t$ limit of the
leading order cross section is about {\it fifty percent smaller} than the
exact leading-order cross section \cite{Dawson2013}.

To clarify to what extent the large radiative corrections observed in the
large-$M_t$ limit can be trusted, we decided to compute additional terms in the 
$1/M_t$ expansion at next-to-leading order in perturbative QCD and to explore
how these power corrections affect the size of next-to-leading order
(NLO) QCD effects.  We note that it
was also found from comparisons of cross sections at leading order
\cite{Dawson2013}, that accounting for additional terms in $1/M_t$ expansion
does not  improve the agreement between expanded and exact results
for moderate and high values of the partonic center-of-mass energy
  $\sqrt{s}$.  In this situation, knowing additional terms in the expansion
may not help with the phenomenology {\it per se}, but it is still useful for
understanding the uncertainty that needs to be assigned to the NLO QCD
prediction for $pp \to HH$.  Moreover, if many terms in $1/M_t$ expansion are
available at next-to-leading order, we may try to improve on the quality of
the expansion by factoring out the exact leading order cross section. This
strategy is known to work very well in the case of single Higgs production in gluon
fusion, giving us a reason to believe that it will be useful for the
process $pp \to HH$ as well.  With the $1/M_t$ expansion at hand, we will be able 
to determine the range of partonic center-of-mass energies, for which this procedure 
works  well.

The remainder of this paper is organized as follows.  We explain  the computational methods in  Section~\ref{comp} and 
describe numerical results  in Section~\ref{numres}.   We present our conclusions in Section~\ref{conc}. 

\section{ Computational methods}
\label{comp}

In this Section we present technical details of  our  computation of the NLO QCD corrections to  
Higgs boson pair production 
in proton collisions, $pp \to HH$. The cross section 
is  given by the product of parton distribution functions and the partonic cross section for 
the process $ij \to HH$, where 
$i$ and $j$ are the colliding partons
\begin{equation} 
\sigma(pp \to HH +X) = \sum \limits_{ij} \int {\rm d} x_1 {\rm d} x_2  f_i(x_1) f_j(x_2) \sigma_{ij \to HH}.
\end{equation} 
At leading order in the strong coupling constant,  both partons are gluons; at next-to-leading order, 
quark-gluon and quark-antiquark  collisions start to contribute.

\begin{figure}[t]
  \begin{center}
    \includegraphics[width=0.6\columnwidth]{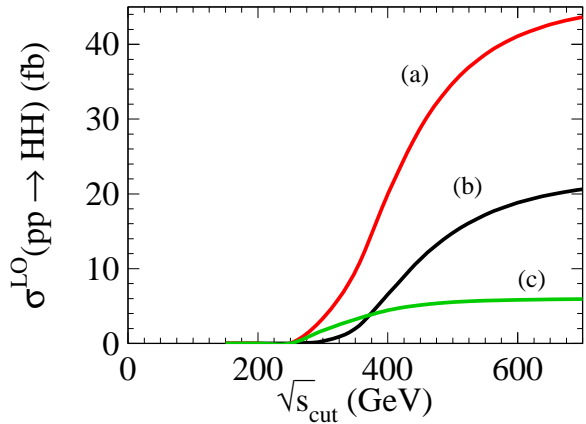}
    \caption{Leading order hadronic cross section for  Higgs boson pair production at the 
    $14~{\rm TeV}$ LHC 
    as the function of the upper cut on the Higgs boson pair invariant mass. 
    Curve (b) is the full result; curve (a) is the box contribution;
    curve (c) is the 
    triangle contribution.  The destructive interference between box and triangle contributions is apparent. 
    We use MSTW2008 parton distribution functions \cite{mstw}.  }
    \label{figm1}
  \end{center}
\end{figure}

We take the gluon fusion channel $gg \to HH$ as an example to describe computational methods. At leading order, this process 
occurs through a box diagram $gg \to HH$ or  through a triangle diagram $gg \to H^*$ with subsequent decay $H^* \to HH$, 
see Fig.~\ref{fig0}.
Both the box and the triangle diagrams are mediated by the top quark loops.
Assuming   that the top quark mass $M_t$ is much larger than both, the mass of the 
Higgs boson $m_H$  and the collision energy of the two gluons, 
 we can understand the  leading contribution to $gg \to HH$ in this large-$M_t$ limit using the concept  
of the effective Lagrangian. The effective Lagrangian 
 for Higgs-gluon interaction can be obtained  by integrating out the top quark field from the Standard Model Lagrangian. 
  It reads \cite{Dawson:1998py}
\be
{\cal L} = \frac{\alpha_s}{6\pi} {\rm Tr} \left [ G_{\mu \nu} G^{\mu \nu}\right ] \log \left ( 1+ \frac{H}{v} \right ), 
\label{eq25}
\ee
where $G_{\mu \nu} = G_{\mu \nu}^{a} t^a$ is the gluon field-strength tensor and $v$ is the vacuum expectation value of the Higgs field. 
Expanding Eq.~(\ref{eq25}) to second 
order in $H/v$, we obtain terms that lead to  $gg H$ and $ggHH$ interactions.  
The scattering amplitude  for $g(p_1) g(p_2) \to HH $, that follows from the effective Lagrangian 
Eq.~(\ref{eq25}), reads 
\be
{\cal A}_{\lambda_1 \lambda_2}  = \frac{\alpha_s \delta_{ab}}{8 \pi v} \delta_{\lambda_1 \lambda_2} 
\left [ 
-\frac{4}{3} + \frac{4 m_H^2}{s-m_H^2}
\right ],
\label{eq1}
\ee where $s = 2p_1 \cdot p_2$, $\lambda_{1,2}$ are helicities of the two
gluons and $a,b$ are their color indices.  We note that the first term in
square brackets in Eq.~(\ref{eq1}) comes from the box diagram and the second
term from the triangle diagram, that contains a triple Higgs coupling.  
It is interesting to note that box and triangle contributions to $gg \to HH$ amplitude  tend to 
strongly cancel each other.  For example, at the partonic threshold $s = 4 m_H^2$, the
cancellation between the two contributions is exact. To further illustrate this point,  in Fig.~\ref{figm1} 
we show the leading order Higgs boson pair  production cross section at the $14~{\rm TeV}$ 
LHC in dependence of the upper cut on the partonic center-of-mass collision energy.  As can be seen 
from that Figure, the impact of the triple Higgs boson coupling on the observable 
cross section comes from the destructive interference of the box and triangle contributions. It  
reduces the cross section by almost a factor of two for a broad range of  $\sqrt{s_{\rm cut}}$
and implies  linear sensitivity of   
the $pp \to HH$ production cross section to small changes in triple Higgs boson 
coupling.  It also follows from Fig.~\ref{figm1}  that the $pp \to HH$  cross section gets saturated 
at values of the Higgs pair invariant mass cut $\sqrt{s_{\rm cut}} \approx 600 \sim 700~{\rm GeV}$. 
Therefore, for a reliable NLO computation, we require  accurate  description of partonic cross sections 
for double Higgs boson production up to $\sqrt{s} \sim 600~{\rm GeV}$ since higher 
partonic center-of-mass energies are not important for the hadronic $pp \to HH$ cross section.

While Eq.~(\ref{eq1}) is useful for thinking about Higgs boson production, it
does not lead to an adequate description of $pp \to HH$ cross section since
further terms in $1/M_t$ expansion are needed.  Deriving such terms is
straightforward at leading order in perturbative QCD where one can simply
obtain an exact result by computing the required one-loop
diagrams. Unfortunately, it becomes prohibitively difficult to do that at
next-to-leading order where computation of two-loop box diagrams for $gg \to
HH$ with full $M_t$-dependence is required.  To overcome this problem, we
expand the relevant NLO cross sections in powers of $1/M_t$.  We do so by
employing a procedure that has been used for the computation of finite-$M_t$
corrections to the production cross section of a single Higgs boson in gluon
fusion \cite{Pak:2009dg}.  In what follows we briefly describe it.

\begin{figure}[t]
\begin{center}
  \mbox{}\quad\quad\quad\includegraphics[width=0.95\columnwidth]{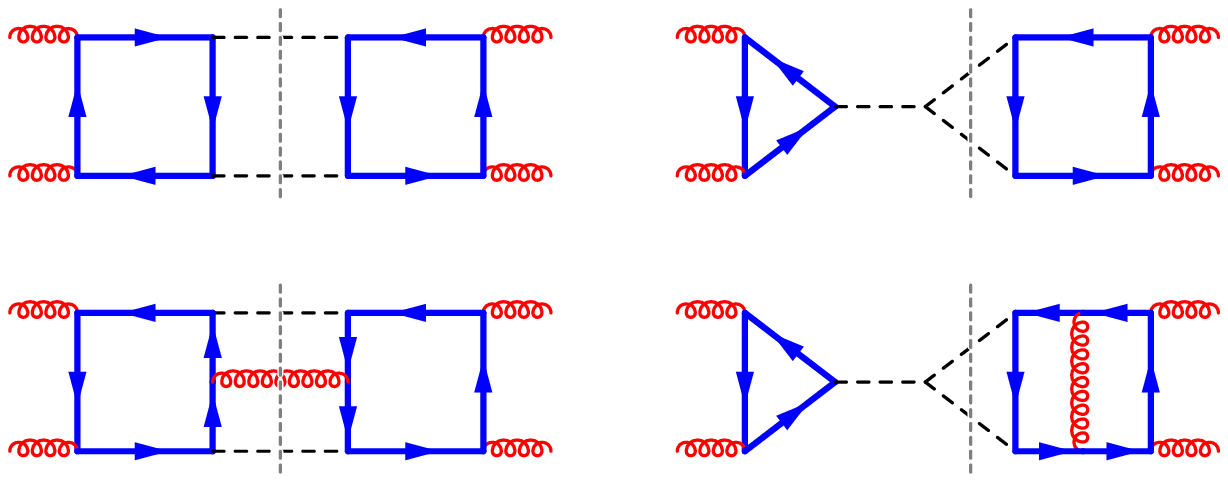}
  \\[3em]
  \includegraphics[width=0.95\columnwidth]{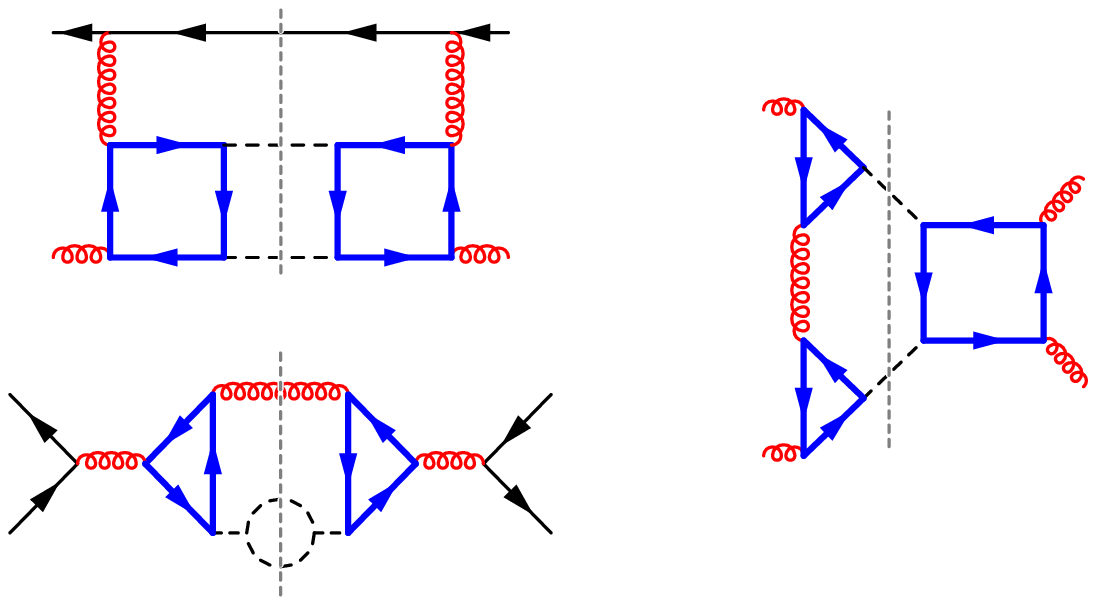}
  \\[2em]
  \caption{Examples of forward scattering amplitudes that we need to
    consider. Dashed vertical lines represent unitarity cuts. Solid lines
    are top  and light quarks,  dashed  lines are Higgs bosons.}
\label{fig0a}
\end{center}
\end{figure}

We start by representing the total cross section for the double Higgs boson
production by  the properly-normalized  
discontinuity of the gluon forward scattering amplitudes (see Fig.~\ref{fig0a} 
for sample Feynman diagrams),
\be
\sigma( gg \to HH) \sim {\rm Disc} \left [ {\cal A}_{gg \to gg} \right ]
\,.
\label{eq11}
\ee The discontinuity is obtained by cutting diagrams that contribute to
forward-scattering amplitudes through two Higgs lines at leading and
next-to-leading order, and through two Higgs lines and a gluon line at
next-to-leading order.  As explained in Ref.~\cite{am}, Eq.~(\ref{eq11}) is
helpful because the procedure of taking the discontinuity of the forward
scattering amplitude ${\cal A}_{gg \to gg} $ commutes with modern techniques
that can be used to compute it.  Indeed, to calculate diagrams that contribute
to ${\cal A}_{gg \to gg}$, we first perform the large-$M_t$ expansion by
treating masses and momenta of the Higgs bosons and, where appropriate, of a
cut gluon line, to be much smaller than the top quark mass.  Once this
expansion is carried out, we obtain a large number of Feynman integrals and we
reduce them to a minimal set of master integrals using the
integration-by-parts technique \cite{Tkachov:1981wb,Chetyrkin:1981qh}.  The
number of master integrals is small -- we require one integral for leading and
seven integrals for next-to-leading order computation.  The calculation proceeds
through a sequence of computer-algebra programs developed for computing mass
corrections to single Higgs boson production in gluon fusion
\cite{Pak:2009dg}.  Relevant three- and four-loop diagrams that contribute to
the forward scattering amplitude are obtained with {\sf QGRAF} \cite{qgraf},
supplemented with additional scripts to remove unnecessary diagrams.  Each
diagram is then expanded assuming $M_t^2 \gg m_H^2$ and $M_t^2 \gg s$ using
programs {\sf q2e} and {\sf exp} \cite{q2eexp}.  We express each Feynman
diagram as linear combination of scalar integrals using {\tt
  FORM}~\cite{Vermaseren:2000nd} and its parallel version {\tt
  TFORM}~\cite{Tentyukov:2007mu}.  The reduction to master integrals is
performed using {\tt FIRE}~\cite{Smirnov:2008iw,Smirnov:2013dia}\footnote{We
  thank A.V.~Smirnov and V.A.~Smirnov for allowing us to use the unpublished
  {\tt C++} version of {\tt FIRE}.}  which implements the Laporta
algorithm~\cite{laporta}. The first two expansion terms in $\rho$ are computed
for general QCD gauge parameter $\xi$ and the independence of the final result
of $\xi$ is used as a check of the computation. Furthermore, our result for
the $\rho^0$ contribution to the partonic cross section agrees with
Ref.~\cite{Dawson:1998py}.

\begin{figure}[t]
  \begin{center}
    \includegraphics[width=0.95\columnwidth]{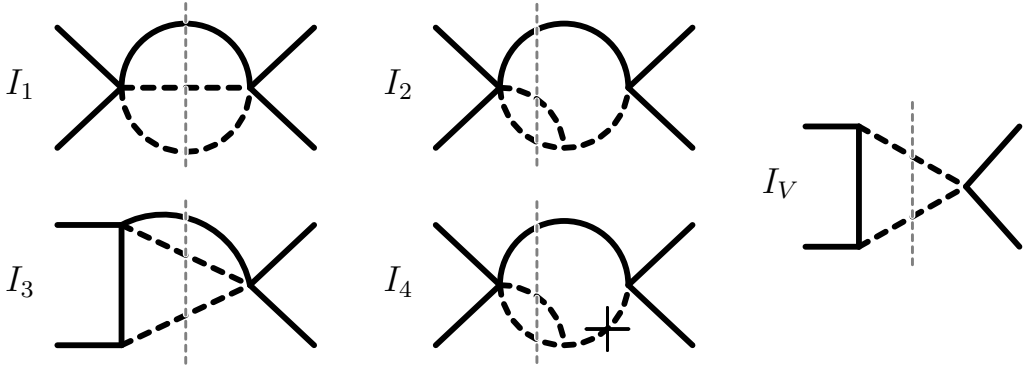}
    \caption{The master integrals $I_1$, $I_2$, $I_3$, $I_4$ and $I_V$.
      Dashed lines cut through propagators that are replaced by the mass-shell
      conditions.  Dashed lines are the Higgs boson propagators and solid
      lines correspond to massless scalar propagators. In the
      case of $I_4$ the cross indicates a propagator raised to power minus
      one.  $I_V$ contributes to the virtual corrections at next-to-leading
      order.}
    \label{fig1}
  \end{center}
\end{figure}

Sample master integrals that appear at NLO are shown in Fig.~\ref{fig1}.  We
will discuss their computation in what follows. Note that we require these
master integrals to higher orders in the dimensional-regularization parameter
$\epsilon = (d-4)/2$ since they multiply the divergent reduction coefficients.
In principle, it is possible to derive exact expressions for these integrals
but this is cumbersome.  However, it is very easy to construct an expansion of
these integrals around the double Higgs partonic threshold to, essentially,
arbitrary order in the expansion parameter.  We now show how this is done for
individual master integrals.

For the leading order process $g(p_1) + g(p_2) \to H(p_3) +H(p_4)$,
there is just one master integral which corresponds to the available
phase space for final state particles \be I_0 =  \int [{\rm d} p_3
] [{\rm d} p_4] (2\pi)^d \delta^{(d)}(p_1 + p_2 - p_3 - p_4 ), \ee where
$[{\rm d} p] = {\rm d}^{d-1}p/((2\pi)^{(d-1)} 2 p_0)$, $p_1^2 = p_2^2 = 0$ and $p_3^2 = p_4^2 = m_H^2$.
At next-to-leading order, several different types of master integrals appear
since ${\cal O}(\alpha_s)$ corrections can either come from real gluon
emissions or from virtual gluon exchanges. For the real emission corrections,
four master integrals appear as the result of the reduction but only three of
them are linearly independent. For the virtual corrections, we need two-loop
vacuum bubble integrals and one additional master integral that we describe
below.

The four  master integrals that we require to compute for the real emission corrections are 
(see Fig.~\ref{fig1})
\be
\begin{split} 
& I_1 = \int  [{\rm d} p_3 ] [{\rm d} p_4] [{\rm d} p_5] 
(2\pi)^{d}\delta^{(d)}(p_1 + p_2 - p_3 - p_4 - p_5) , \\
& I_2 = \int  [{\rm d} p_3 ] [{\rm d} p_4] [{\rm d} p_5] 
(2\pi)^{d}\delta^{(d)}(p_1 + p_2 - p_3 - p_4 - p_5)  \times ( (p_3+p_4)^2 - m_H^2), \\
& I_3 = \int  [{\rm d} p_3 ] [{\rm d} p_4] [{\rm d} p_5] 
(2\pi)^{d}\delta^{(d)}(p_1 + p_2 - p_3 - p_4 - p_5)  \times ( p_2 - p_4)^2, \\
& I_4 = \int  [{\rm d} p_3 ] [{\rm d} p_4] [{\rm d} p_5] 
(2\pi)^{d}\delta^{(d)}(p_1 + p_2 - p_3 - p_4 - p_5)  \times ((p_3+p_4)^2 - m_H^2)^{-1}.
\end{split} 
\ee

To compute $I_1$, it is convenient to introduce integration over total
momentum of the two Higgs bosons $Q = p_3 + p_4$ by inserting ${\rm d}^dQ
\delta^{(d)}(Q -p_3 - p_4) = 1 $ into the integrand of $I_1$. We obtain
\be 
  I_1 = \int \limits_{4 m_H^2}^{s} \frac{{\rm d} Q^2}{(2\pi)} {\rm
  Lips}(Q,p_3,p_4) {\rm Lips}(P_{12} , Q, p_5),
  \label{eq7}
\ee
where $P_{12} = p_1 + p_2$ and ${\rm Lips}$ is the two-particle
Lorentz-invariant phase space defined as 
\be
{\rm Lips}(p,q,r) = \int [{\rm d} q] [{\rm d}  r] (2\pi)^d \delta^{(d)} ( p - q - r).
\ee
It is straightforward to compute the phase spaces that appear in  Eq.~(\ref{eq7}). We find 
\be
\begin{split}
&  {\rm Lips}(Q,p_3,p_4) =  Q^{-2 \epsilon} \frac{\Omega_{d-1}
  2^{2\epsilon}}{(2\pi)^{d-2} 8}
\left ( 1 - \frac{4m_H^2}{Q^2} \right )^{1/2-\epsilon}, \\
& { \rm Lips}(P_{12},Q,p_5) =  s^{- \epsilon}
\frac{\Omega_{d-1} 2^{2\epsilon}}{(2\pi)^{d-1} 8} \left ( 1 - \frac{Q^2}{s} \right )^{1-2\ep},
\end{split}
\ee
where $s = P_{12}^2$ and $\Omega_{d} = 2\pi^{d/2}/\Gamma(d/2)$ is the solid angle of the $d$-dimensional space.
Combining the above formulas, we obtain  
\be
I_1 = {\cal N} s^{-\ep} 
\int \limits_{4 m_H^2}^{s} {\rm d} Q^2 Q^{-2\ep} \left ( 1 - \frac{4 m_H^2}{Q^2} \right )^{1/2-\ep} 
\left ( 1 - \frac{Q^2}{s} \right )^{1-2\ep},
\label{eqi1q}
\ee
where the normalization factor ${\cal N}$ reads 
\be
{\cal N} = \left [ \frac{\Gamma(1+\ep)}{(4\pi)^{d/2}} \right ]^2 
\left [ 1 + 4 \epsilon + \left ( 12 - \frac{2\pi^2}{3} \right ) \ep^2 
+ {\cal O}(\ep^3) 
\right ].
\ee
To facilitate the computation of the integral, we change the integration variable by writing 
$Q^2 = s ( 1- \delta \mu)$ and express the result in terms of the variable $\delta = 1 - 4m_H^2/s$.
We obtain 
\be
I_1 = {\cal N} s^{1-2\ep} \delta^{5/2-3\ep}
\int \limits_{0}^{1} 
\frac{{\rm d}\mu}{\sqrt{1-\delta \mu}} (1-\mu)^{1/2-\ep} \mu^{1-2\ep}.
\label{eqi12}
\ee
The integrand of $I_1$ can now be expanded in a Taylor series in $\delta$ and the resulting 
integrals can be evaluated in a straightforward way. 

We now discuss the evaluation of $I_2$. Note that the difference between $I_2$
and $I_1$ is due to additional terms in the integrand that, however, only
depend on $Q^2$ since $(p_3+p_4)^2 - m_H^2 = Q^2 - m_H^2$. Therefore, the
integral representation for $I_2$ can be easily deduced from
Eq.~(\ref{eqi1q}). We find 
\be I_2 = {\cal N} s^{-\ep} \int \limits_{4
  m_H^2}^{s} {\rm d} Q^2 Q^{-2\ep} \left ( 1 - \frac{4 m_H^2}{Q^2} \right
)^{1/2-\ep} \left ( 1 - \frac{Q^2}{s} \right )^{1-2\ep} \times (Q^2 - m_H^2).
\label{eqi2q}
\ee
Trading  $Q^2$ for $\mu$, we obtain 
\be
I_2 = {\cal N} s^{2-2\ep} \delta^{5/2-3\ep} 
\int \limits_{0}^{1} 
\frac{{\rm d} \mu}{\sqrt{1-\delta \mu}} (1 - \mu)^{1/2-\epsilon} \mu^{1-2\ep} 
\left [ \frac{3}{4} + \delta \left ( \frac{1}{4} - \mu \right ) \right ],
\ee
and the Taylor expansion in $\delta$ becomes straightforward. 

As the next step, we show that the integral $I_3$ is a linear combination of $I_1$ and $I_2$.  Indeed, since 
\be
 I_3 =  \int  [{\rm d} p_3 ] [{\rm d} p_4] [{\rm d} p_5] 
(2\pi)^{d} \delta^{(d)}(p_1 + p_2 - p_3 - p_4 - p_5)  ( m_H^2 - 2p_2 \cdot p_4 ),
 \ee
we can simplify it by using the following equations
\be
\int [{\rm d} p_3] [{\rm d} p_4 ]  (2\pi)^{d}\delta^{(d)}(Q - p_3 - p_4)  ( m_H^2 - 2 p_2 \cdot p_4) 
= ( m_H^2 - p_2 \cdot Q ) {\rm Lips}(Q,p_3,p_4),
\ee
and 
\be
\begin{split}
& \int [{\rm d} p_5 ][{\rm d} Q] (2\pi)^{d}\delta^{(d)}( P_{12} - p_5 - Q) ( m_H^2 - p_2 \cdot Q)\\
&= \int [{\rm d} p_5 ][{\rm d} Q] (2\pi)^{d}\delta^{(d)}( P_{12} -
p_5 - Q) \left ( m_H^2 - \frac{p_2 \cdot P_{12} ( s+Q^2)}{2s} \right )
\\  
&= \left ( m_H^2 - \frac{s+Q^2}{4}  \right ) {\rm Lips}(P_{12},Q,p_5) .
\end{split} 
\ee
Therefore, $I_3$ becomes 
\be
I_3= {\cal N} s^{-\ep} 
\int \limits_{4m_H^2}^{Q^2} 
{\rm d}Q^2 Q^{-2\ep} 
\left ( 1 - \frac{4m_H^2}{Q^2} \right )^{1/2-\ep} 
\left ( 1 - \frac{Q^2}{s} \right )^{1-2\ep} 
\left ( m_H^2 - \frac{s+Q^2}{4} \right ).
\ee
Using Eqs.~(\ref{eqi1q}) and~(\ref{eqi2q}),  it is easy to see that $I_3$ can be represented as a linear combination of $I_1$ and 
$I_2$
\be
I_3 = \left ( m_H^2 - \frac{s}{4} \right ) I_1 - \frac{1}{4} (I_2 + m_H^2 I_1 ).
\ee

Finally, the integrand for $I_4$ differs from that for 
$I_2$ because the term $(Q^2-m_H^2)$ occurs in
the denominator instead of the numerator; 
therefore, the useful representation for $I_4$ can be easily derived following
the steps described above. We obtain
\be
I_4 = 4{\cal N} s^{-2\ep} \delta^{5/2-3\ep} 
\int \limits_{0}^{1} \frac{{\rm d} \mu}{\sqrt{1-\delta \mu}}
\frac{(1-\mu)^{1/2-\ep} \mu^{1-2\ep} 
}{3 + \delta (1 - 4\mu) },
\ee
and we can expand it in $\delta$ in a straightforward way.

Finally, additional master integrals are needed for the virtual corrections. The virtual corrections required for this  
are the two-loop ones and their large-mass expansion leads to two distinct contributions. The first contribution arises 
when the loop momenta are comparable to the mass of the top quark. In this case, the two-loop diagrams are 
expandable in external momenta and the Higgs boson masses.  Using the integration-by-parts, 
the two-loop integrals are mapped onto a single two-loop vacuum bubble integral.  In a situation where  one loop-momentum 
is comparable to the top mass and the other loop-momentum is comparable to the Higgs boson mass or 
to the partonic collision energy, a two-loop diagram factorizes 
into a product of one-loop diagrams and we do not discuss it here. 
In addition,  at next-to-leading order  diagrams appear where two Higgs bosons are produced in a way that involves 
exchange of a gluon in a $t$-channel, see Fig.~\ref{fig0a}.
The computation of these diagrams requires a new  master integral. It reads
(see Fig.~\ref{fig1})
\be
I_V = \int [{\rm d} p_3] [{\rm d} p_4]  (2\pi)^d \delta(p_1+p_2 - p_3 - p_4) \times (p_1 - p_3)^{-2}.
\ee
We calculate it following the above discussion and  obtain 
\be
I_V = - {\cal N}_V 8 \pi s^{-1-\ep}\delta^{1/2-\ep} 
\int \limits_{0}^{1} 
{\rm d} x \frac{x^{-\ep} (1-x)^{-\ep} (1+\delta) }{ (1+ \delta)^2 - 4 \delta (1-2 x)^2 },
\ee
where
\be
{\cal N}_V =  \frac{\Gamma(1+\ep)}{(4\pi)^{d/2}} 
\left ( 1 - \frac{\pi^2}{6} \ep^2 + {\cal O}(\ep^3) 
\right ) . 
\ee
Expanding $I_V$  in $\delta$ is straightforward.

Once the reduction of the contributing diagrams to master integrals is performed
and once explicit expressions for master integrals are substituted, we obtain
unrenormalized results for partonic cross sections.  To obtain physical
results, we need to renormalize the strong coupling constant and the top quark
mass to remove the ultraviolet divergences, and to renormalize parton
distribution functions to remove the collinear singularities associated with
gluon emissions from the initial state.  For the ultraviolet renormalization,
we use the one-loop expressions
\be
\begin{split}
  \alpha_s^{\rm bare}  = \frac{\mu^{2\epsilon} \alpha_s}{S_\epsilon} 
  \left ( 1 - \frac{\beta_0}{\epsilon} \frac{\alpha_s}{2\pi}  + \cdots \right ),
  \;\;\;\;
  M_{t}^{\rm bare } =  M_t 
  \left [ 1 - \frac{C_F \alpha_s}{\pi S_\epsilon} 
    \left( \frac{3}{4 \epsilon} - 1 - \frac{3}{4}\ln\frac{\mu^2}{M_t^2}\right) + \cdots \right ].
  \label{eq::ren}
\end{split} 
\ee Here, $S_\epsilon = (4\pi)^{-\ep} e^{\gamma \epsilon}, \gamma =
0.5772\ldots$ is the Euler constant, $\alpha_s \equiv \alpha_s(\mu) $ is the
$\overline {\rm MS}$ QCD coupling constant evaluated at the scale $\mu$,
$\beta_0 = 11 N_c/6 - n_f/3$ is the one-loop QCD beta-function and $M_t$ is
the pole top quark mass.  For consistent ultraviolet renormalization, 
we  first use $n_f=6$  but express our final result for cross section 
through $\alpha_s$ with five
active flavours by using the one-loop decoupling relation for the top
quark. Ellipses in Eq.~(\ref{eq::ren}) stand for contributions suppressed by additional powers of
$\alpha_s$.

The collinear renormalization is performed in the standard way, by redefining
parton distribution functions.  The corresponding modification of the NLO
cross sections, required to make it finite, reads \be \sigma_{gg}^{(1)}(s) =
\delta {\bar \sigma}_{gg}^{(1)}(s) + \frac{1}{\ep}  \int \limits_{x_{\rm
    min}}^{1} {\rm d} x P_{gg}(x) \; \sigma_{gg}^{(0)} (x s), \ee where
$x_{\rm min} = 4 m_H^2/s$ is determined from the Higgs pair production
threshold in $gg$ collisions and $P_{gg}$ is the gluon splitting function 
\be
P_{gg} = C_A \left [ \frac{1}{(1-x)_+} -1 + \frac{1-x}{x} + x(1-x) \right ]
+ \delta(1-x) \left ( \frac{11 C_A}{12} - \frac{n_f}{6} \right )\,, 
\ee 
with $C_A=3$ for QCD.  Similar equations hold also for $qg$, $\bar q g$ and
$q \bar q$ production channels; since they are standard, we do not present
them here.

\section{Numerical results} 
\label{numres} 

In this Section we present our results for the NLO QCD corrections to the
double Higgs boson production cross section at the LHC.  Values of the Higgs
boson mass and the on-shell top quark mass are taken to be $m_H
= 126~{\rm GeV}$~\cite{atlasd,cmsd} and $M_t = 173.18~{\rm
  GeV}$~\cite{Aaltonen:2012ra}, respectively.  We 
begin with discussing partonic cross sections at leading and next-to-leading
order.  It is convenient to express the cross sections using two variables, $x
= 4 m_H^2/s = 1-\delta$ and $\rho = m_H^2/M_t^2$.  As we explained
in the previous Section, we compute the $HH$ production cross section as
series in $\rho = m_H^2/M_t^2$. 
 The expansion starts at $\rho^{0}$ and we are able to obtain
five  terms of the $\rho$-expansion, up to ${\cal O}(\rho^4)$, for the $gg$
partonic channel and seven terms, up to ${\cal O}(\rho^6)$, for the $qg$ and $q
\bar q$ channel.  The master integrals are computed as an expansion in the
parameter $\delta = 1-x$; for the final results all terms up to ${\cal O}(\delta^{50})$
are included. 
The partonic cross sections are defined as
\be
\sigma_{ij \to H+X}(s,\rho) = 
 \delta_{ig} \delta_{jg} \sigma_{gg}^{(0)}(s,\rho) + \frac{\alpha_s}{\pi}\sigma^{(1)}_{ij}(s,\rho),
\ee
where $\sigma^{(1)}_{ij}$ is the ${\cal O}(\alpha_s)$ correction to the leading 
order cross section.
For the discussion of the partonic cross sections,  we set the factorization 
and the renormalization scales to $ \mu_r = \mu_f = 2 m_H$. We will describe the scale 
dependence of our results below when we consider the hadronic cross section.  

\begin{figure}[t]
  \begin{center}
    \includegraphics[width=0.48\columnwidth]{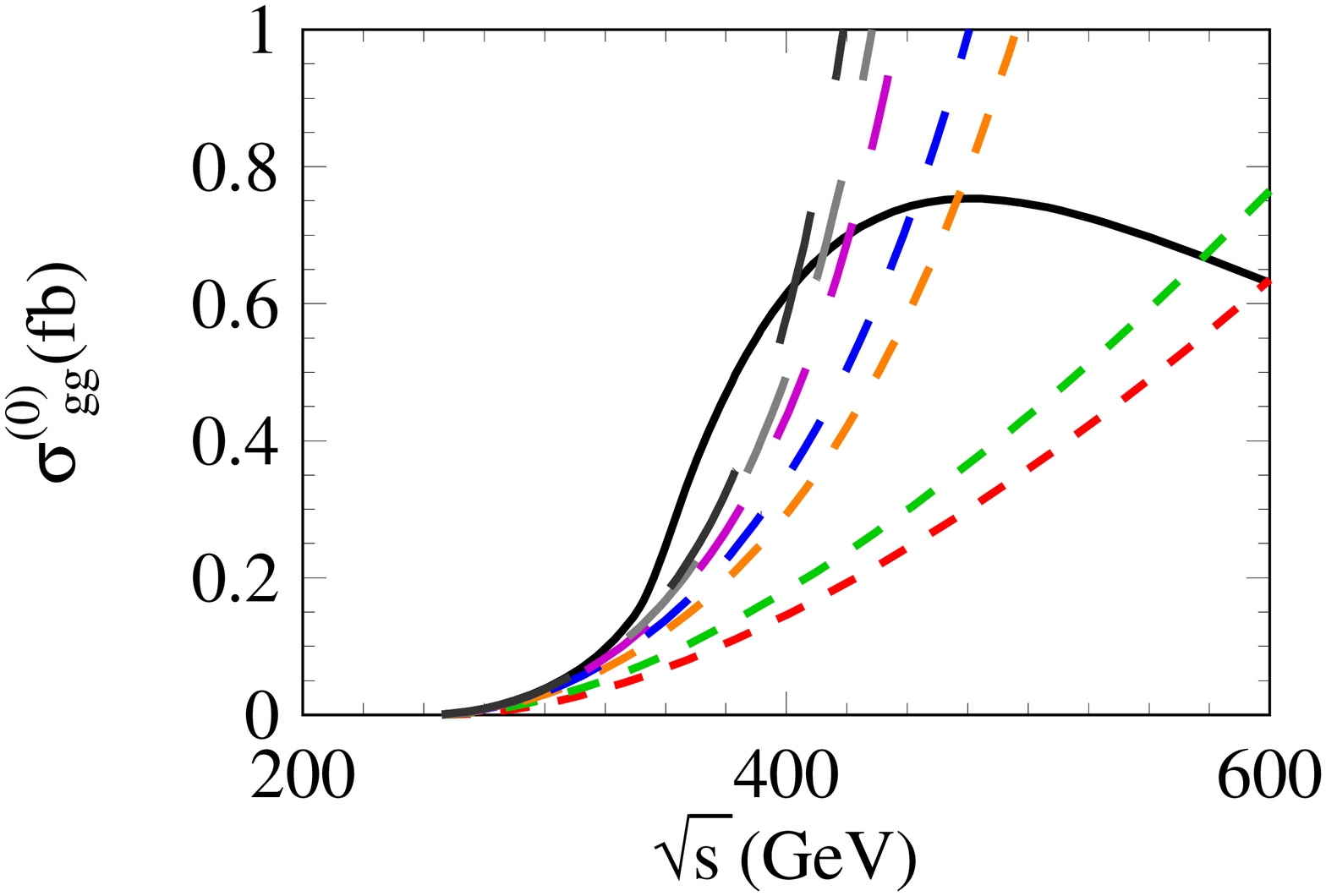}
    \includegraphics[width=0.48\columnwidth]{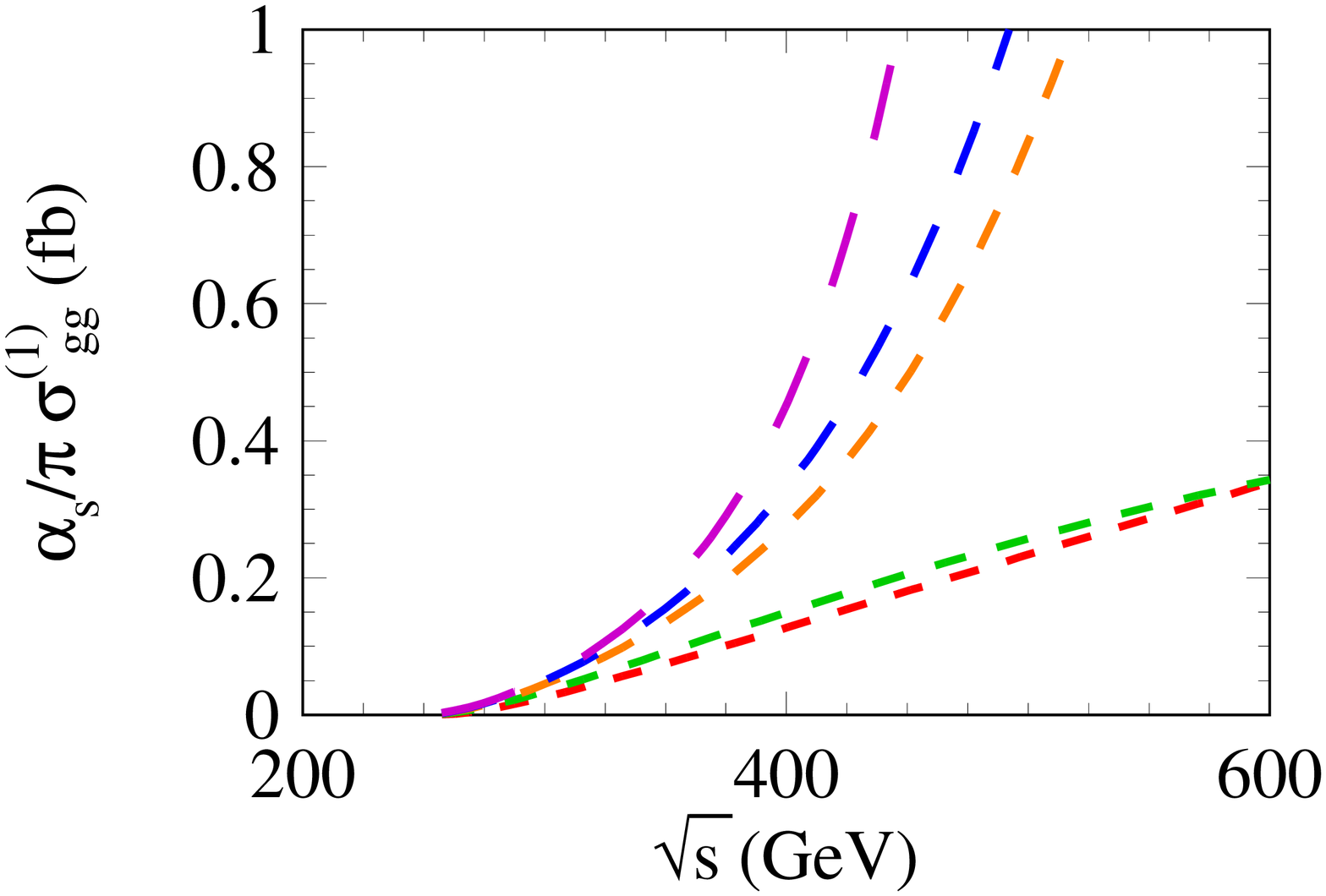}
    \caption{Leading order partonic $gg \to HH$ cross section (left pane) and
      next-to-leading order {\it contribution} to $gg \to HH$ cross section
      $\alpha_s/\pi \times \sigma_{gg}^{(1)}$ (right pane), in
      fb. Different lines correspond to 1) exact leading order cross section
      -- black solid; 2) cross sections expanded 
      to ${\cal O}(\rho^0)$ -- short-dashed red; 
      to ${\cal O}(\rho^1)$ -- short-dashed green; 
      to ${\cal O}(\rho^2)$ -- dashed orange; 
      to ${\cal O}(\rho^3)$ -- dashed blue;
      to ${\cal O}(\rho^4)$ -- dashed violet;
      to ${\cal O}(\rho^5)$ -- long-dashed light gray;
      to ${\cal O}(\rho^6)$ -- long-dashed dark gray;
      See text for the description of
      input parameters. }
\label{fig3}
\end{center}
\end{figure}

We begin by showing some  results for the $gg$  channel.  In Fig.~\ref{fig3} we compare 
$\sigma_{gg}^{(0)}(s,\rho)$  with seven approximate cross sections   that are obtained  by expanding $\sigma_{gg}^{(0)}$ 
in $\rho$ through ${\cal O}(\rho^{i})$, $i=0,\ldots,6$.
It is clear from Fig.~\ref{fig3} that  the convergence 
of the expansion is poor. Indeed, already at $\sqrt{s} \sim  350~{\rm GeV}$,  
there is a sizable difference  between the exact and the expanded result.  
In the right pane of Fig.~\ref{fig3} we show the NLO contribution 
to the cross section expanded to different orders in $\rho$.   Similar to the leading order case, 
the $1/M_t$ expansion does not appear to converge. 

\begin{figure}[t]
  \begin{center}
    \includegraphics[width=0.48\columnwidth]{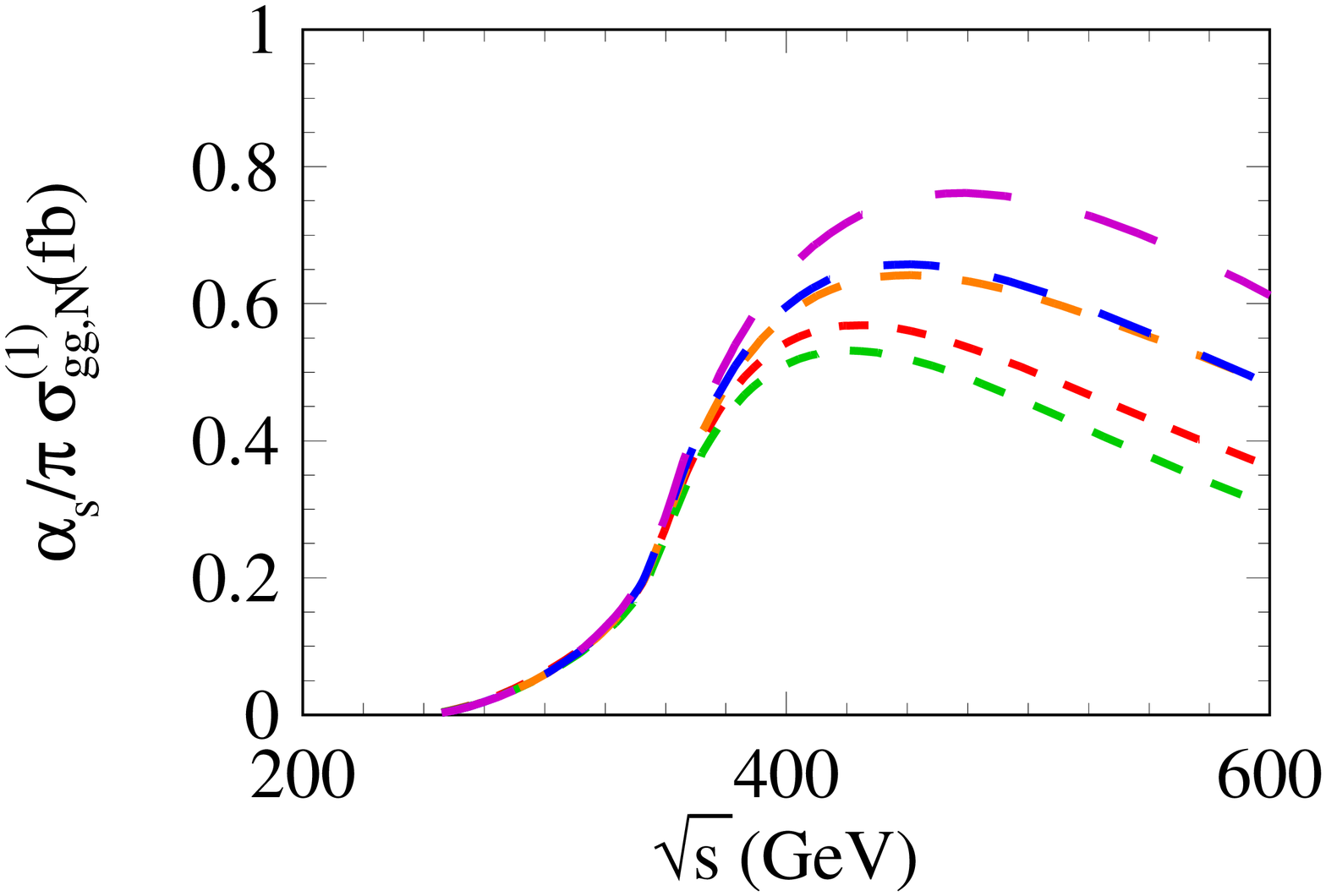}
    \includegraphics[width=0.48\columnwidth]{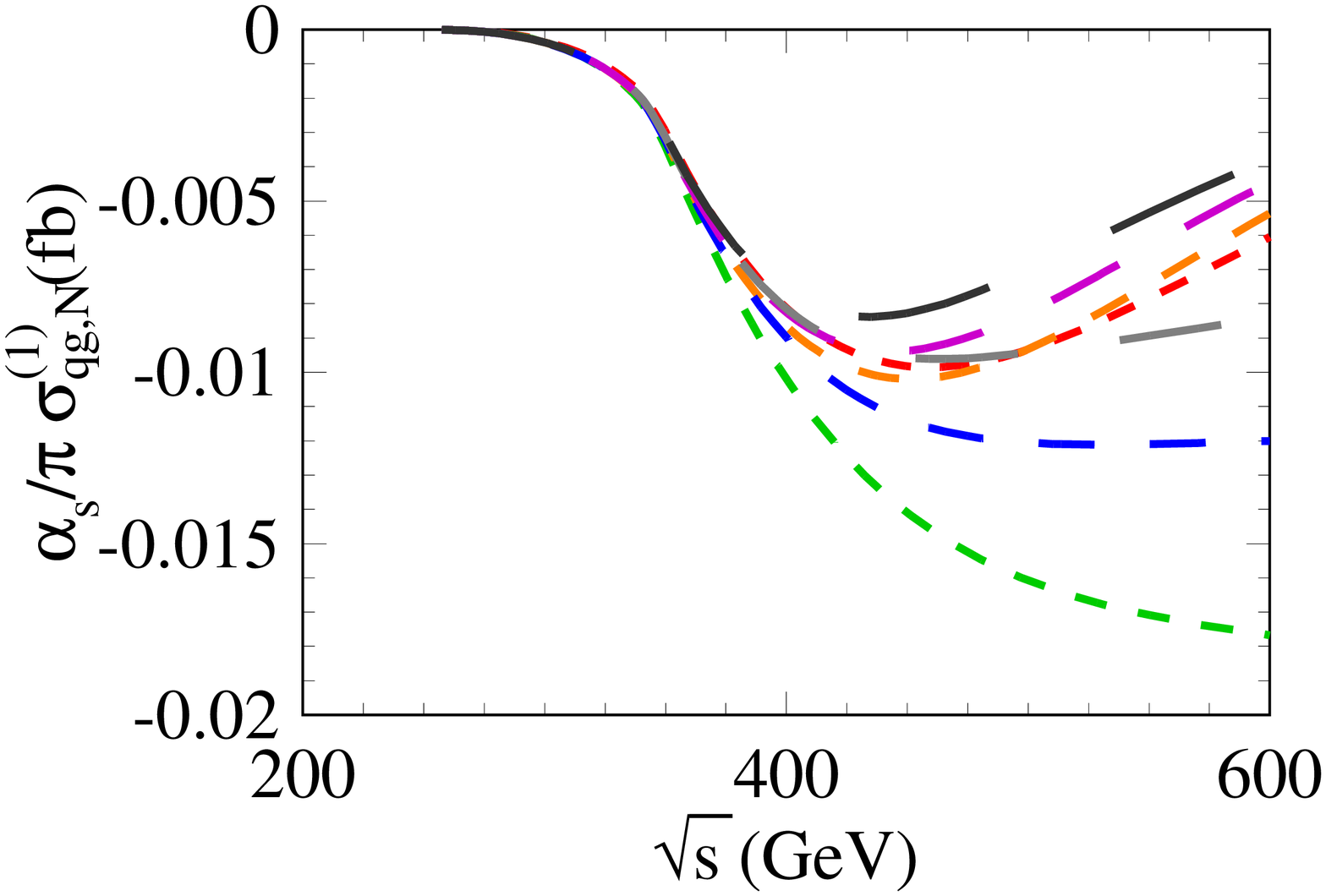}
    \\
    \includegraphics[width=0.48\columnwidth]{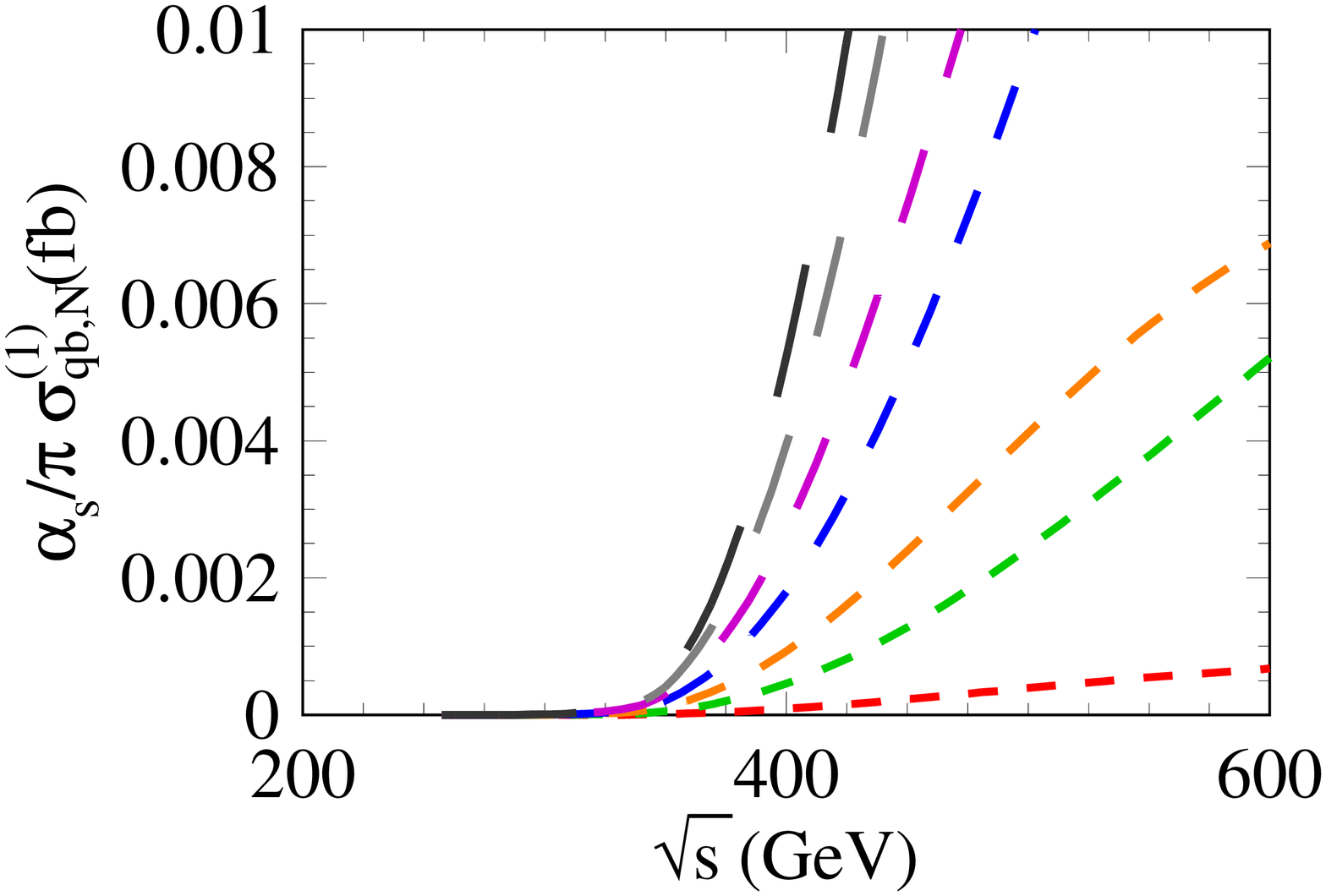}
    \caption{ Next-to-leading order contribution to $gg \to HH$, $qg \to HH$
      and $q \bar q \to HH$ cross sections re-scaled by exact leading order
      result, in fb. The color coding is as in Fig.~\ref{fig3}.}
    \label{fig4}
  \end{center}
\end{figure}

The bad convergence of the $1/M_t$ expansion should not be very
surprising. Indeed, we note that the expansion is not supposed to work beyond,
or even close to, the top quark threshold that occurs at $\sqrt{s} = 2 M_t$. 
Therefore, using the expansion techniques described above, we can only hope to
obtain reliable results for values of $s \ge 4 M_t^2$ if we can show that
corrections do not strongly depend on $s$.  From this perspective, the
situation is similar to what occurs in the single Higgs boson production in
gluon fusion where the applicability of radiative corrections computed in the
large-$M_t$ approximation is usually extended by combining them with the {\it exact}
leading order cross section for $gg \to H$.  The validity of such an approach
in single Higgs production is verified by comparing it to the exact results at
NLO \cite{hnloexact} and by its consistency with known power corrections to
the large-$M_t$ limit at NNLO QCD
\cite{powerc1,powerc2,Pak:2009dg,Pak:2011hs,powerc4}.  Motivated by the
success of this approach to QCD corrections in single Higgs boson production,
we apply it to Higgs pair production as well.  We write the NLO QCD
contribution to the partonic cross section as 
\be
\label{eqnlo}
\sigma_{ij,N}^{(1)} = 
\sigma_{gg,\rm exact}^{(0)}  \Delta_{ij}^{(N)},
\;\;\;\;\;  \Delta_{ij}^{(N)} = 
\frac{\sigma_{ij,\rm exp}^{(1)}}{\sigma_{gg,\rm exp}^{(0)}} = 
\frac{\sum \limits_{n=0}^{N} c_{ij,n}^{\rm NLO}\rho^n}{\sum \limits_{n=0}^{N} c_{gg,n}^{\rm LO}\rho^n} ,
\ee
where both numerator and denominator of the $\Delta$-factor 
are expanded to the same order in $\rho$. By changing $N$ in 
the above formula, we can check the stability of our computation against  additional power corrections. 
Ideally, $\Delta_{ij}^{(N)}$,  should become $N$-independent, after sufficient number of terms are included in the numerator and 
denominator in Eq.~(\ref{eqnlo}).

\begin{figure}[t]
  \begin{center}
    \includegraphics[width=0.48\columnwidth]{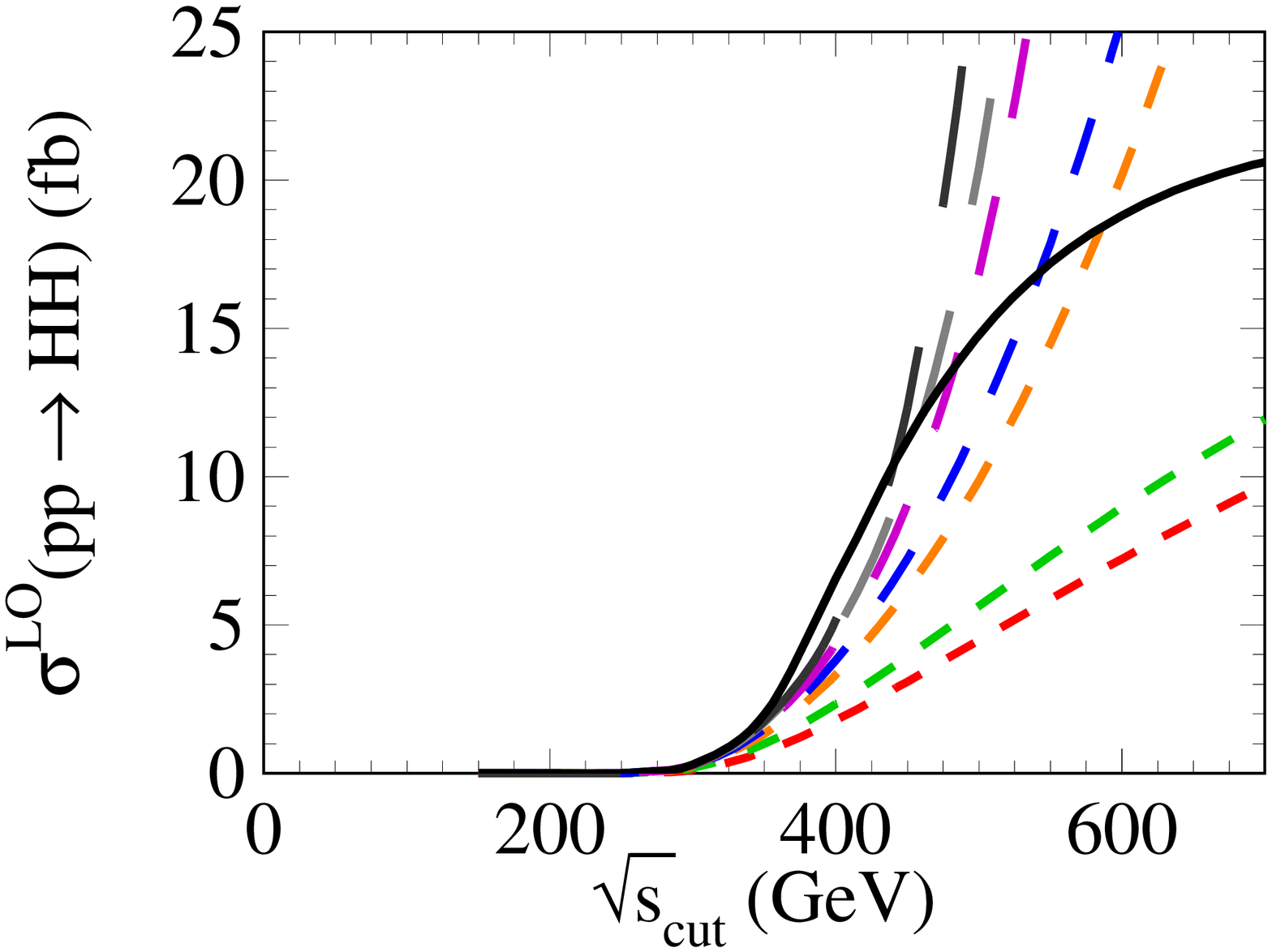}
    \includegraphics[width=0.48\columnwidth]{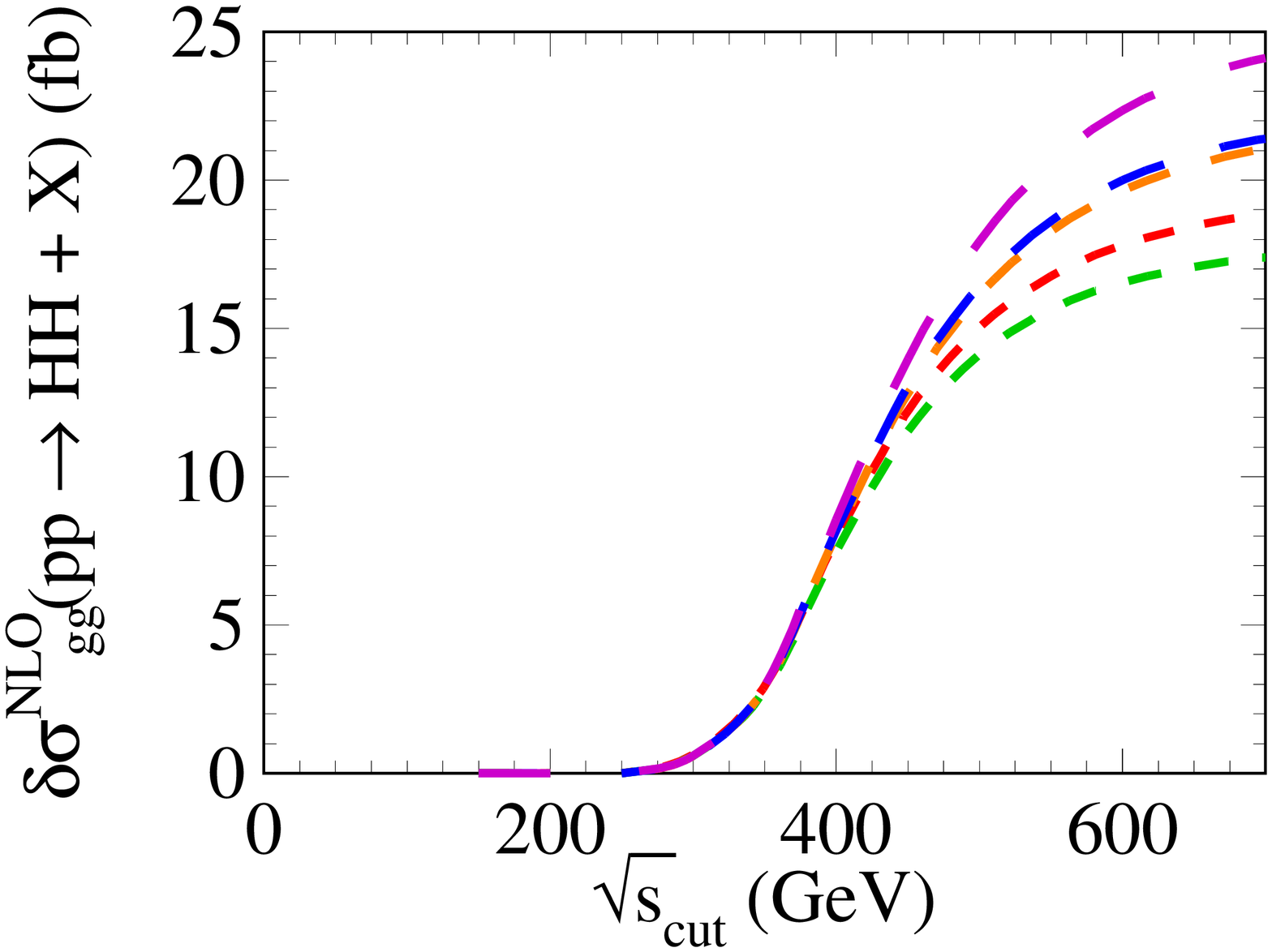}
    \\
    \includegraphics[width=0.48\columnwidth]{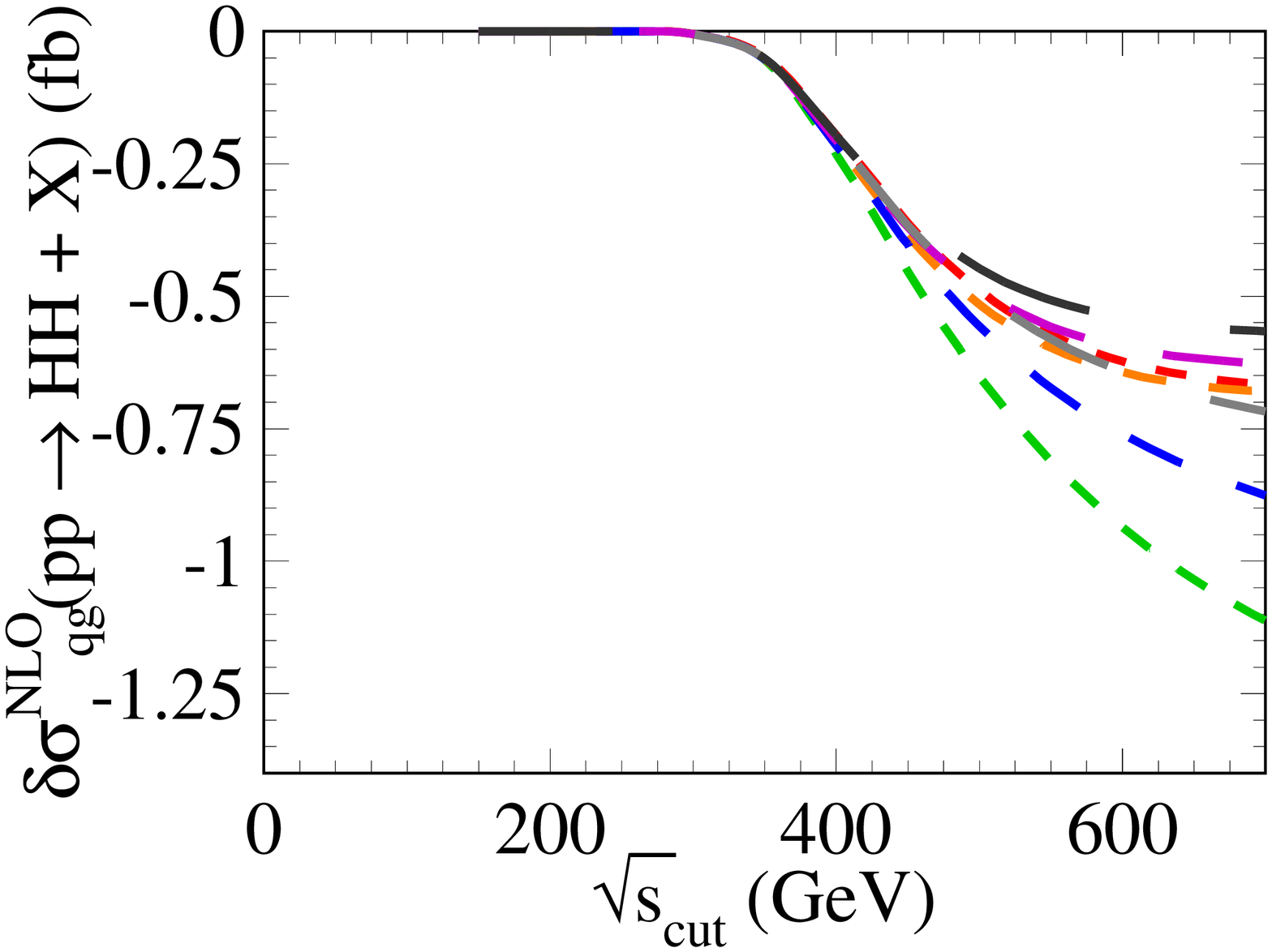}
    \includegraphics[width=0.48\columnwidth]{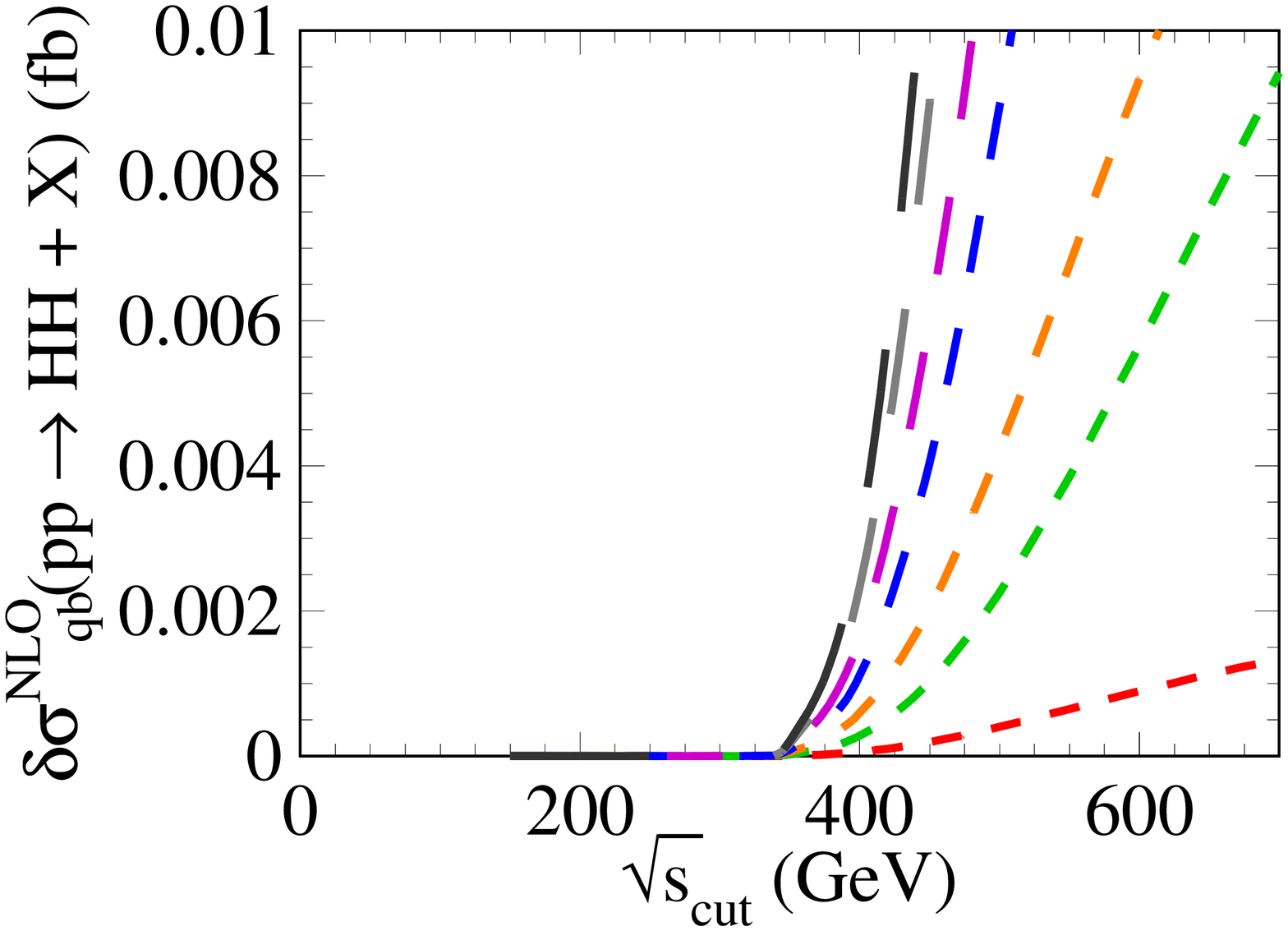}
    \caption{ Leading and next-to-leading order contributions to hadronic
      production cross sections of Higgs boson pairs. Different curves
      represent  the cross section obtained through expansion to different
      orders in $\rho$; rescaling by the leading order cross section is always
      included. The color coding is as in Fig.~\ref{fig3}.}
    \label{fig5}
  \end{center}
\end{figure}

Our results for $\sigma_{gg,N}^{(1)}$ are shown in Fig.~\ref{fig4}. We see
that for $N=0,1,2,3,4$, $\sigma_{gg,N}^{(1)}$ is fairly stable over a broad
range of partonic center-of-mass energies.  In fact, there is practically no
difference between $\sigma_{gg,2}^{(1)}$ and $\sigma_{gg,3}^{(1)}$ all the way
up to $\sqrt{s} \sim 600~{\rm GeV}$ but, unfortunately, $\sigma_{gg,4}^{(1)}$
shows a $13$ percent increase relative to $\sigma_{gg,3}^{(1)}$ for $\sqrt{s}
\sim 450~{\rm GeV}$.  The situation is similar for the $qg$ channel and it is
much worse for $q \bar q$ channel where even with as many as seven terms of
the $1/M_t$ expansion the convergence seems poor, see Fig.~\ref{fig4}.  To
understand the differences between partonic channels, we note that the leading
order cross section $\sigma_{gg}^{(0)}$ does not properly describe the
kinematic features of $\sigma_{qg}^{(1)}$ and $\sigma_{q \bar q}^{(1)}$.
Indeed, the $qg $ scattering, for example, mainly occurs through a $t$-channel
gluon exchange which, for large $s$, has little to do with the
leading order process of a gluon fusion into a pair of Higgs bosons.
Fortunately, these subtleties do not impact predictions for hadronic cross
sections because the $qg$ and $q \bar q$ channels are numerically small --
they provide at most ten percent correction to the next-to-leading order
contribution to Higgs boson pair production cross section. Therefore, even
relatively low accuracy for $q g$ and $q \bar q$ channels is sufficient for
reliable phenomenology.

We proceed now to the discussion of hadronic cross sections.  To obtain numerical results  shown below we employ   
MSTW2008  parton distribution functions (PDFs) \cite{mstw}.  
We consistently use leading order PDFs to compute leading order cross sections 
and next-to-leading order PDFs to compute next-to-leading order cross sections.   We assume the energy of the LHC 
to be $14~{\rm TeV}$.   Values of the strong coupling constants
$\alpha_s(M_{\rm Z})$ that we use 
in our computation are obtained from the MSTW PDF fit. Their leading and next-to-leading order values are  
$0.139384$ and $0.120176$, respectively.  

We  present  results for hadronic cross sections 
as a function of the upper cut on the partonic center-of-mass energy $\sqrt{s_{\rm cut}}$.  
This allows us to explore the stability of the  $1/M_t$ expansion of the Higgs boson pair 
production cross section as we move from the threshold region, where this expansion works
well, to the high-energy region where this expansion becomes less reliable. 
We can also use this cut  as a proxy for the  Higgs boson pair invariant mass 
 cut  which  may be useful  for enhancing observability of  the triple Higgs  boson coupling.  We note 
that at leading order in QCD, a cut on partonic center of mass energy 
is equivalent to the cut on the invariant mass of the Higgs boson pair while at next-to-leading order  the equivalence is not exact anymore. 
For all results below, we use the NLO contributions to partonic cross sections
re-scaled with exact leading order cross section, cf. Eq.~(\ref{eqnlo}).

Our results for leading and next-to-leading order contributions to hadronic
cross sections are shown in Fig.~\ref{fig5}.  The NLO QCD predictions for
hadronic cross sections appear to be more reliable than similar results for
partonic cross sections; the reason for this is the known enhancement of gluon
luminosity at small values of $s$ which reduces contributions of large-$s$
partonic processes to hadronic cross sections.  To illustrate this point, we
note that even for $\sqrt{s_{\rm cut}} \sim 700~{\rm GeV}$, the shift in
$\delta \sigma_{gg}^{\rm NLO}$ due to the last computed term in the $1/M_t$
expansion is close to twelve percent. For lower values of $\sqrt{s_{\rm cut}}$
the convergence is significantly better; for $\sqrt{s_{\rm cut}} \sim 450~{\rm
  GeV}$, a similar shift is close to seven percent.  We note that these numbers
refer to NLO contributions to next-to-leading order cross sections and that the
numerical significance of these shifts is further ameliorated by large leading
contribution.

 \begin{figure}[t]
  \begin{center}
    \includegraphics[width=0.48\columnwidth]{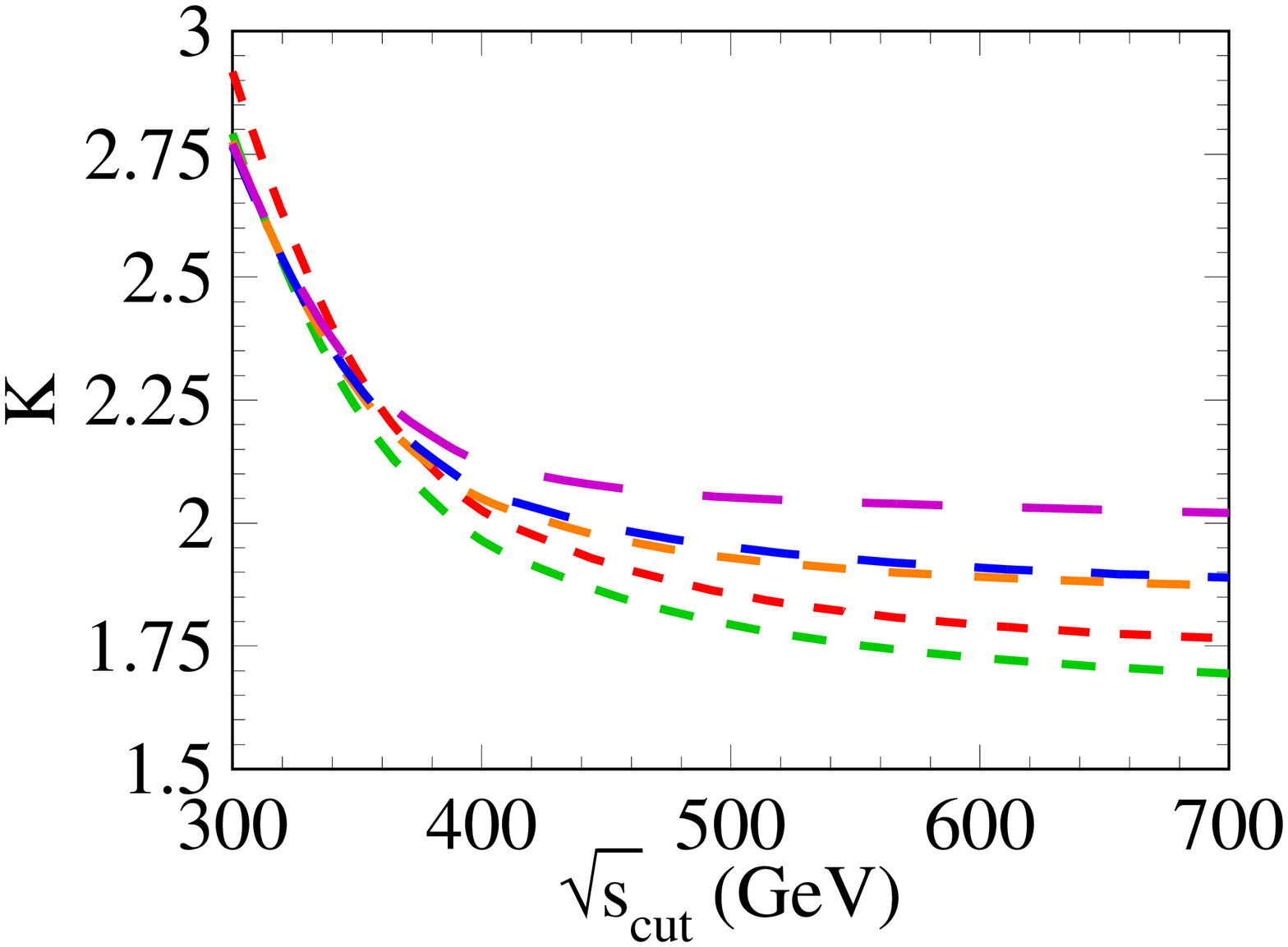}
    \includegraphics[width=0.48\columnwidth]{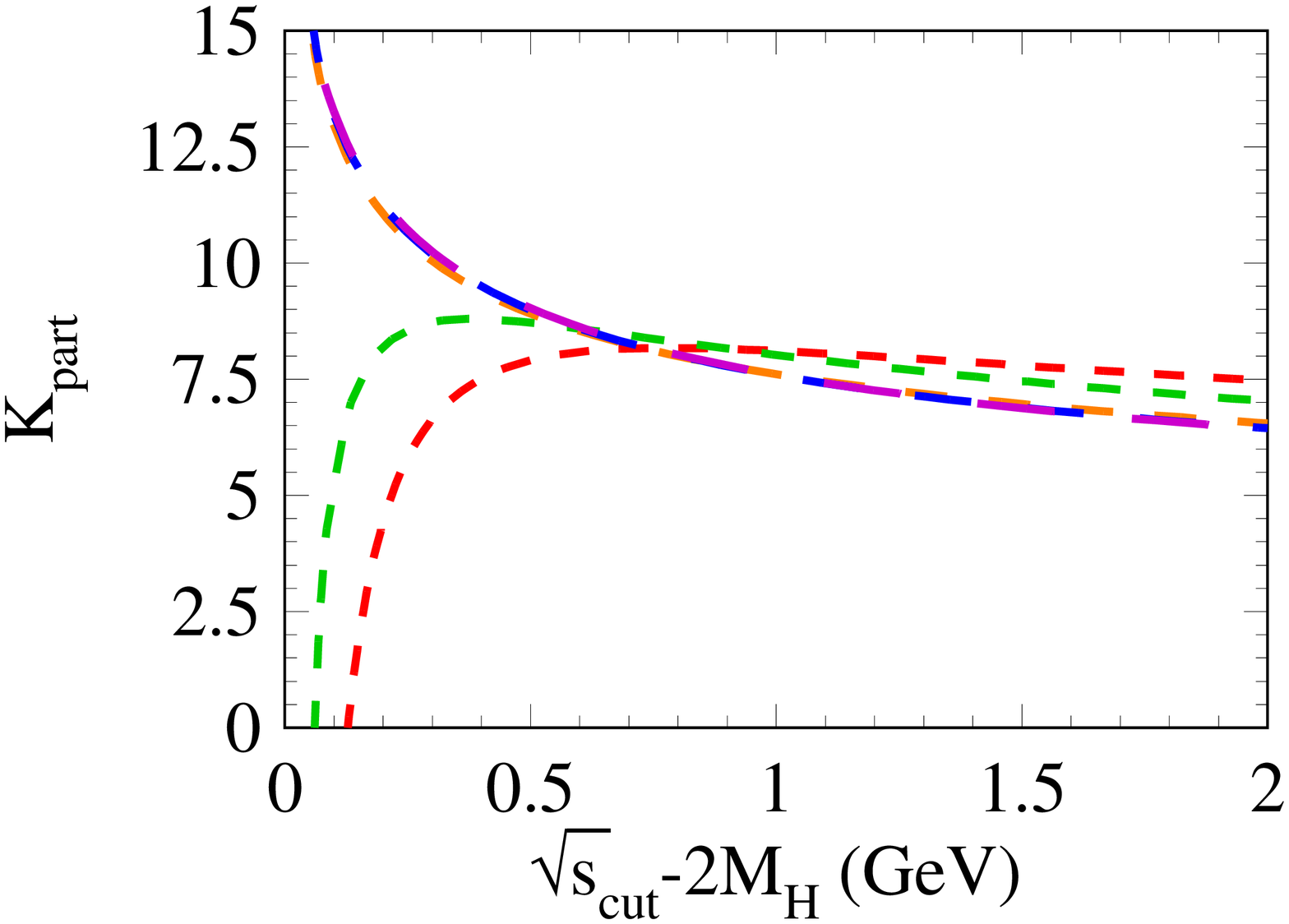}
    \caption{ The NLO $K$-factors defined as the ratio of NLO and LO hadronic
      (left) and partonic (right) cross sections are shown as the function of
      $\sqrt{s_{\rm cut}}$.  The renormalization and factorization
      scales are chosen to be $\mu = 2 m_H$. The color coding is as in
      Fig.~\ref{fig3}.} 
    \label{fig7}
  \end{center}
\end{figure}

As we already mentioned in the Introduction, QCD corrections to  Higgs boson pair
production cross section 
at the LHC are  large. To illustrate this, 
in the left pane of Fig.~\ref{fig7} we show the $K$-factor,  defined as the ratio of NLO and LO  
 $pp \to HH+X$ cross sections, evaluated 
at $\mu = 2m_H$, as a function  of $\sqrt{s_{\rm cut}}$.  
In general, the $K$-factors are large, confirming earlier 
observation of Ref.~\cite{Dawson:1998py}.  However, the $K$-factor  is also 
strongly dependent on $s_{\rm cut}$,  decreasing  significantly  from 
the threshold region $\sqrt{s_{\rm cut}} \approx 2 m_H$ to large values of $s_{\rm cut}$.  
This change of the $K$-factor 
is related to additional suppression of the leading order 
cross section for $gg \to HH$ in the threshold region that we 
mentioned in the Introduction and the fact that for the NLO amplitude such suppression  is not present.
Because of that, the $K$-factor in the threshold 
region is enhanced.  To illustrate this point, we show the partonic $gg$ cross section expanded to 
first non-vanishing orders in $\rho$ and $\delta$ at leading and next-to-leading order in the threshold 
region\footnote{The ${\cal O}(\rho^0)$ leading order partonic cross section scales as $\delta^{5/2}$
at threshold.}
\begin{align}
  & \sigma_{gg} \approx \frac{G_F^2  m_H^2}{2\pi} 
  \left ( \frac{\alpha_s}{\pi} \right )^2 \left \{  
      \frac{7 \rho }{25920} \delta^{3/2} 
      + \rho^2 \left( 
        \frac{49}{4147200}\delta^{1/2}
        + \frac{24001}{87091200}\delta^{3/2}
      \right)
    \right.
    \nonumber\\&\mbox{}
    \left.
    + 
    \frac{\alpha_s}{\pi}
    \left [ 
      - \frac{\delta^{3/2}}{1296}  
      + \rho \left ( 
        -\frac{7}{103680} \sqrt{\delta}  
        + \delta^{3/2} \left\{ \frac{8957}{466560} 
          - \frac{7 \pi^2}{25920} 
          + \frac{7}{1620}\lmmh
          + \left (-\frac{7}{270} 
        \right.\right. \right. \right. \right. 
        \nonumber\\ & \left. \left. \left. \left. \left. 
            - \frac{7}{2160} \ln \frac{\mu^2}{m_H^2} 
          \right  ) \ln 2 + \frac{7}{540} \ln^2 2 
          + \left  (- \frac{7}{810} 
            + \frac{7}{720} \ln 2  
            - \frac{7}{4320} \ln \frac{\mu^2}{m_H^2}
          \right ) L_\delta  +
          \frac{7}{4320}  L_\delta^2 
        \right \}
      \right  )
      \right.\right.  \nonumber\\&\mbox{} \left.\left.
      + \rho^2 \left(
        \left\{
          \frac{128399}{261273600}
          - \frac{49}{4147200} \pi^2
          + \frac{49}{345600}\lmmh
          + \frac{49}{86400}\ln^2 2
          + \frac{49}{691200} L_\delta^2
        \right.\right.\right.\right.  \nonumber\\&\mbox{} \left.\left. \left.\left.   
          + \left(
            - \frac{49}{57600}
            - \frac{49}{345600}\lmmh
          \right)\ln 2 
          + \left(
            - \frac{49}{172800}
            - \frac{49}{691200}\lmmh
            + \frac{49}{115200}\ln 2
            \right) L_\delta
        \right\}\delta^{1/2}
        \right.\right.\right.  \nonumber\\&\mbox{} \left.\left.\left.          
        +\left\{
          \frac{318919}{15676416}
          - \frac{24001}{87091200} \pi^2
          + \frac{193037}{43545600}\lmmh
          + \frac{24001}{1814400}\ln^2 2
        \right.\right.\right.\right.  \nonumber\\&\mbox{} \left.\left. \left.\left.   
          + \frac{24001}{14515200} L_\delta^2
          + \left(
            - \frac{24001}{907200}
            - \frac{24001}{7257600}\lmmh
          \right)\ln 2 
        \right.\right.\right.\right.  \nonumber\\&\mbox{} \left.\left. \left.\left.   
          + \left(
            - \frac{382987}{43545600}
            - \frac{24001}{14515200}\lmmh
            + \frac{24001}{2419200}\ln 2
            \right) L_\delta
        \right\}\delta^{3/2}
      \right)
    \right ] + \ldots \right \}
\end{align} 
where $L_\delta = \ln \delta$ and ellipses stand for terms additionally
suppressed by powers of either $\rho$ or $\delta$.  We observe that in the
limit $\delta \to 0$, dominant contributions to $\sigma_{gg}$ come from a term
${\cal O}(\alpha_s^2 \rho^2 \sqrt{\delta})$ at leading order and from a term $
{\cal O}(\alpha_s^3 \rho^2 \sqrt{\delta} \ln \delta^2)$ at next-to-leading
order.  This implies that the behavior of the $K$-factor in the $\delta \to 0$
limit is strongly affected by power-suppressed $1/M_t$ terms and that the
$K$-factor becomes infinite at the exact $\delta \to 0$ threshold for $HH$
production.  This point is further illustrated in the right pane of
Fig.~\ref{fig7}, where {\it partonic} $K$-factors are shown in the vicinity of
the two-Higgs threshold.  It follows from that plot that the threshold limit
of the ${\cal O}(\rho^{0,1})$ curves is significantly different from the
approximation that includes ${\cal O}(\rho^2)$ and higher-power corrections.
We also note that $1/M_t$-corrections change the hadronic $K$-factor by about
$14$ percent at $\sqrt{s_{\rm cut}} = 700~{\rm GeV}$ and the change decreases
for smaller values of $s_{\rm cut}$.  The shift in the $\sqrt{s_{\rm cut}} =
700~{\rm GeV}$ $K$-factor due to the last computed $1/M_t$ correction is close
to seven percent.

\begin{figure}[t]
  \begin{center}
    \includegraphics[width=0.5\columnwidth]{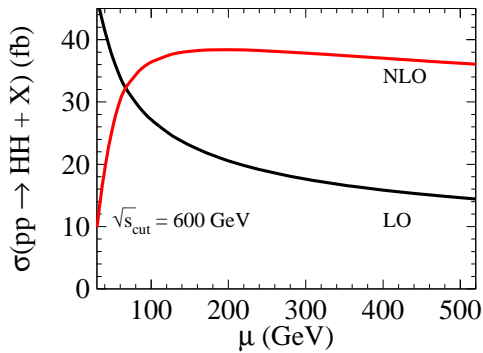}
    \caption{ Scale dependence of the hadronic production cross section for
      $pp \to HH$.  }
    \label{fig6}
  \end{center}
\end{figure}

In Fig.~\ref{fig6}, we show the residual dependence of the production cross
section $pp \to HH$ on the factorization and renormalization scales that we
set equal to each other.  The NLO cross section is computed with all available
$1/M_t$ corrections included.  The cut on the partonic center-of-mass
collision energy of $600~{\rm GeV}$ is imposed.  It follows from
Fig.~\ref{fig6} that the NLO QCD cross section is practically independent of
the renormalization and factorization scales in a broad range of $\mu$.
Choosing $\mu = 2m_H$ as the central value and estimating the
uncertainty by increasing and decreasing $\mu$ by a factor of two, we arrive
at the NLO cross section estimate $\sigma_{pp \to HH} = 38^{+0}_{-2}~{\rm fb}$
for $\sqrt{s_{\rm cut}} = 600~{\rm GeV}$ to be compared to $\sigma_{pp \to HH}
= 18^{+6}_{-4}~{\rm fb}$ at LO. The scale-dependence uncertainty of the NLO
cross section is therefore close to five percent, a significant improvement
compared to ${\cal O}(30\%)$ uncertainty of the leading order cross section.
 
\begin{figure}[t]
  \begin{center}
    \includegraphics[width=0.5\columnwidth]{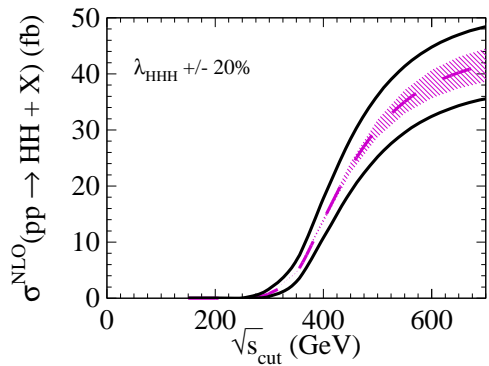}
    \caption{The NLO hadronic cross section at the $14~{\rm TeV}$ LHC as a
      function of $\sqrt{s_{\rm cut}}$.  Two black curves correspond to $\pm
      20\%$ variation in the triple Higgs boson coupling relative to its SM
      value.  The violet (hatched) band shows the uncertainty in the SM prediction for
      $pp \to HH$ due to uncalculated $1/M_t$ corrections.  }
    \label{fig8}
  \end{center}
\end{figure}

Finally, we briefly discuss implications of our results for the extraction of
the triple Higgs boson coupling by comparing variations in $pp \to HH$ cross
section induced by changes in $\lambda_{HHH}$ and the uncertainties in the
Standard Model prediction for that cross section.  In particular, we focus on
the uncertainties caused by imperfect knowledge of the $1/M_t$ corrections to
the cross section.  For our best estimate of the $pp \to HH$ cross section, we
use exact leading order result and next-to-leading contributions expanded to
highest known power in $\rho$ (i.e. $\rho^4$ for $gg$ channel and $\rho^6$ for
$qg$ and $q \bar q$ channel).  We assign the one-sided error to this cross
section by comparing it with a similar computation that employs NLO cross
sections expanded through $\rho^3$ in $gg$ channel and $\rho^6$ in $qg$ and $q
\bar q$ channels.  In Fig.~\ref{fig8} we show the NLO QCD cross sections for
double Higgs boson production for three values of triple Higgs boson coupling,
as a function of the cut on partonic center-of-mass energy.  Two black curves
correspond to $\pm 20\%$ variations in the triple Higgs boson coupling; such
variations cause ${\cal O}(20\%)$ changes in the cross section since, as we
discussed in the Introduction, the impact of the triple Higgs boson coupling
on the $pp \to HH$ cross section is caused by the destructive interference of
box and triangle contributions.  The violet (hatched) band shows Standard Model (SM)
prediction for $pp \to HH$ including the uncertainty of the $1/M_t$ expansion.
We conclude from Fig.~\ref{fig8} that the quality of our current knowledge of
the Higgs boson pair production cross section, inasmuch as $1/M_t$ corrections
are concerned, is sufficient to detect ${\cal O}(10 \%)$ deviations in Higgs
triple boson coupling, relative to its Standard Model value.
 
\section{Conclusions} 
\label{conc}

In this paper, we studied top quark mass corrections to the production cross
section of the Higgs boson pair at the LHC through next-to-leading order in
perturbative QCD. This is an interesting process since it allows us to directly explore
the self-interaction potential of the Higgs field.
QCD corrections to Higgs boson
pair production are known to be very large \cite{Dawson:1998py}; they enhance
the Higgs pair production cross section by almost a factor two.  For technical
reasons, these corrections were originally computed \cite{Dawson:1998py} in
the infinite top quark mass approximation which, for realistic Higgs boson
masses, has very limited applicability.  For phenomenological applications, it
is important to check the stability of this result against $1/M_t$ power
corrections.  Computation of such mass corrections at next-to-leading order in
QCD is the main goal of this paper.

We calculated power corrections through ${\cal O}(1/M_t^8)$ for $gg \to HH$
partonic channel and through ${\cal O}(1/M_t^{12})$ for $q g$ and $q \bar q$.
These corrections turned out to be large and poorly convergent which is hardly
surprising since similar behavior can be observed already at leading order,
where many terms in $1/M_t$ expansion can be compared to the exact result.  We
have shown that the problem of poor convergence can be cured if the exact
leading-order cross section is used to normalize the NLO QCD corrections.
With such normalization, we find that mass corrections provide ${\cal O}(10\%)$
increase in the NLO QCD prediction for $pp \to HH$ production cross section.

Our computation provides a realistic estimate of the NLO QCD effects in the
total cross section for a Higgs boson pair production.  It justifies the use
of the large-$M_t$ approximation to describe QCD corrections to this process
and opens up a way to reliably estimate NNLO QCD corrections to $pp \to
HH$.\footnote{Very recently,
  next-to-next-to-leading order soft and 
  virtual corrections were computed in
  this limit~\cite{deFlorian:2013uza}.}
Given a very small dependence of the cross section on the renormalization
and factorization scales, which may appear to be accidental given the
magnitude of the NLO QCD corrections, computation of NNLO QCD corrections is
important for quantifying theoretical uncertainty in $\sigma(pp \to HH)$.
Moreover, it is worth remembering that observation of $pp \to HH$ is difficult
and requires good control of kinematic properties of Higgs bosons decay
products. It remains an important and challenging problem to extend NLO QCD
results presented in this paper to describe kinematic distributions relevant
for $pp \to HH$ observation.

\medskip

\noindent
{\bf Acknowledgments}

The research of J.H, J.G. and M.S. is supported by the Deutsche Forschungsgemeinschaft in the
Sonderforschungsbereich Transregio~9 ``Computational Particle Physics''.
The  research of K.M. is partially supported by US NSF under grants PHY-1214000.
The research of K.M. and J.G. are partially supported  by Karlsruhe Institute of Technology through 
is distinguished researcher fellowship program.


\medskip

\noindent

\begin{thebibliography}{99}

%
%

\bibitem{atlasd} G.~Aad {\it et al.} [ATLAS Collaboration], Phys. Lett. B{\bf 716}, 1 (2012).

\bibitem{cmsd}  S. Chatrchyan {\it et al.}, [CMS Collaboration], Phys. Lett. B{\bf 716}, 30 (2012). 

\bibitem{atlasc}  ATLAS Collaboration,  ATLAS-CONF-2013-034 .

\bibitem{bij}  E.W.N.~Glover and J.J. van der Bij, Nucl. Phys. B{\bf 309}, 282 (1988).

\bibitem{plehn}  T.~Plehn, M.Spira and P.M.~Zerwas, Nucl. Phys. B{\bf 479}, 46 (1996).

\bibitem{dj} A.~Djouadi, W.~Kilian,~M.~M\"uhlleitner and P.M.~Zerwas, 
Eur. Phys. J. C{\bf 10}, 45 (1999).

\bibitem{baur} U.~Baur, T.~ Plehn, and D.~Rainwater, Phys. Rev. D{\bf 69}, 053004 (2004). 

\bibitem{papa}
  A.~Papaefstathiou, L.~L.~Yang and J.~Zurita,
  Phys.\ Rev.\ D {\bf 87} (2013) 011301
  [arXiv:1209.1489 [hep-ph]].

\bibitem{dolan} 
  M.~J.~Dolan, C.~Englert and M.~Spannowsky,
  JHEP {\bf 1210} (2012) 112
  [arXiv:1206.5001 [hep-ph]].

\bibitem{dawson} S.~Dawson,  Nucl. Phys.  B{\bf 359}, 283 (1991).

\bibitem{spira1} A.~Djouadi, M.~Spira and P.M.~Zerwas, 
Phys. Lett. B{\bf 70}, 1372 (1991).

\bibitem{hnloexact}
 M.~Spira, A.~Djouadi, D.~Graudenz and P.~M.~Zerwas,
  Nucl.\ Phys.\ B {\bf 453}, 17 (1995).

\bibitem{Dawson:1998py} 
  S.~Dawson, S.~Dittmaier and M.~Spira,
  Phys.\ Rev.\ D {\bf 58}, 115012 (1998)
  [hep-ph/9805244].

\bibitem{Dawson2013} 
  S.~Dawson, E.~Furlan and I.~Lewis,
  Phys.\ Rev.\ D {\bf 87} (2013) 014007
  [arXiv:1210.6663 [hep-ph]].

\bibitem{mstw}  A.D.~Martin, W.J.~Stirling,  R.S.~Thorne and G.~Watt, 
Eur. Phys. J. C{\bf 63}, 189 (2009).

\bibitem{Pak:2009dg} 
  A.~Pak, M.~Rogal and M.~Steinhauser,
  JHEP {\bf 1002}, 025 (2010)
  [arXiv:0911.4662 [hep-ph]].

\bibitem{am}
  C.~Anastasiou and K.~Melnikov,
  Nucl.\ Phys.\ B {\bf 646}, 220 (2002)
  [hep-ph/0207004].

\bibitem{Tkachov:1981wb} 
  F.~V.~Tkachov,
  Phys.\ Lett.\ B {\bf 100}, 65 (1981).

\bibitem{Chetyrkin:1981qh} 
  K.~G.~Chetyrkin and F.~V.~Tkachov,
  Nucl.\ Phys.\ B {\bf 192}, 159 (1981).

\bibitem{qgraf} P. Nogueira, J. Comp. Phys. {\bf 105}, 279  (1993).

\bibitem{q2eexp} R.~Harlander, T.~Seidensticker and M.~Steinhauser, Phys. Lett. B {\bf 426}, 125 (1998);
  T.~Seidensticker, arXiv:hep-ph/9905298.

\bibitem{Vermaseren:2000nd}
  J.~A.~M.~Vermaseren,
  math-ph/0010025.

\bibitem{Tentyukov:2007mu}
  M.~Tentyukov and J.~A.~M.~Vermaseren,
  Comput.\ Phys.\ Commun.\  {\bf 181} (2010) 1419
  [hep-ph/0702279 [HEP-PH]].

\bibitem{Smirnov:2008iw}
  A.~V.~Smirnov,
  JHEP {\bf 0810} (2008) 107
  [arXiv:0807.3243 [hep-ph]].

\bibitem{Smirnov:2013dia}
  A.~V.~Smirnov and V.~A.~Smirnov,
  arXiv:1302.5885 [hep-ph].

\bibitem{laporta} S.~Laporta and E.~Remiddi, Phys. Lett. B {\bf 379}, 283  (1996);
S.~Laporta, Int. J. Mod. Phys. A~{\bf 15}, 5087  (2000).

\bibitem{Aaltonen:2012ra}
  T.~Aaltonen {\it et al.}  [CDF and D0 Collaborations],
  Phys.\ Rev.\ D {\bf 86} (2012) 092003
  [arXiv:1207.1069 [hep-ex]].

 \bibitem{powerc1}
 S.~Marzani, R.~D.~Ball, V.~Del Duca, S.~Forte and A.~Vicini,
  Nucl.\ Phys.\ B {\bf 800}, 127 (2008)
  [arXiv:0801.2544 [hep-ph]];
  Nucl.\ Phys.\ Proc.\ Suppl.\  {\bf 186}, 98 (2009)
  [arXiv:0809.4934 [hep-ph]];

  \bibitem{powerc2}
  R.~V.~Harlander and K.~J.~Ozeren,
  JHEP {\bf 0911}, 088 (2009)
  [arXiv:0909.3420 [hep-ph]];

\bibitem{Pak:2011hs}
  A.~Pak, M.~Rogal and M.~Steinhauser,
  JHEP {\bf 1109} (2011) 088
  [arXiv:1107.3391 [hep-ph]].

 \bibitem{powerc4}
 R.~V.~Harlander, H.~Mantler, S.~Marzani and K.~J.~Ozeren,
  Eur.\ Phys.\ J.\ C {\bf 66}, 359 (2010)
  [arXiv:0912.2104 [hep-ph]].

\bibitem{deFlorian:2013uza}
  D.~de Florian and J.~Mazzitelli,
  arXiv:1305.5206 [hep-ph].


\end{thebibliography}
\end{document}